\documentclass[aps,twocolumn,a4paper,showpacs]{revtex4}
\usepackage[colorlinks=true, pdfstartview=FitV, linkcolor=blue, citecolor=red, urlcolor=magenta,breaklinks=true]{hyperref}
\usepackage{graphicx}
\usepackage[all]{xy}
\usepackage{amsmath}
\usepackage{amssymb}
\usepackage{color}
\usepackage{subeqnarray}
\usepackage{subfigure}
\usepackage{epstopdf}

\newcommand{\bes}{\begin{subequations}}
\newcommand{\ees}{\end{subequations}}
\def\ben{\begin{eqnarray}}
\def\een{\end{eqnarray}}
\newcommand{\bens}{\begin{subeqnarray}}
\newcommand{\eens}{\end{subeqnarray}}
\def\be{\begin{equation}}
\def\ee{\end{equation}}
\def\e{\text{e}}

\def\exp{\text{exp}}
\def\erf{\text{erf}}
\begin{document}
\title{Hybrid Bloch Brane} 
\author{D. Bazeia} 
\author{Elisama E.M. Lima}
\author{L. Losano}  
\affiliation{Departamento de F\'\i sica Universidade Federal da Para\'iba, 58051-900 Jo\~ao Pessoa PB, Brazil}
\date{\today}
\begin{abstract}
This work reports on models described by two real scalar fields coupled with gravity in the five-dimensional spacetime, with a warped geometry involving one infinite extra dimension. Through a mechanism that smoothly changes a thick brane into a hybrid brane, one investigates the appearance of hybrid branes hosting internal structure, characterized by the splitting on the energy density and the volcano potential, induced by the parameter which controls interactions between the two scalar fields. In particular, we investigate distinct symmetric and asymmetric hybrid brane scenarios.
\end{abstract}
\maketitle


\section{Introduction}

In the last decades, a huge amount of effort has been made to understand problems involving the cosmological constant and hierarchy \cite{CCP,CD,Rubakov,RA,RA2,Arkani,FD,Grem,Csaki}. In this sense, the study of branes in higher dimensional theories became important because it provides a procedure for resolving such questions \cite{Rubakov,RA,Arkani}. Ever since,  interest in exploring the physics of extra dimensions has been growing and growing. In Ref. \cite{RA2}, one introduced a thin braneworld concept which deals with a warped geometry and engenders an infinity extra dimension. Later, this scenario was modified to support thick branes, through the inclusion of background scalar fields coupled to five-dimensional gravity, as investigated in \cite{FD,Grem,Csaki}, for instance. In the absence of gravity, the scalar source field supports kinklike structures which are responsible for the appearance of the thick brane. 

However, thick branes can also have internal structure \cite{BF} and can be generated by two-kink solutions \cite{BB,Dutra}, possessing the advantage of bringing room for the presence of nontrivial structures inside the brane. It provides a richer treatment for the system under consideration, since it allows the manifestation of branes hosting internal structure. In special, in Ref. \cite{BB} the authors suggested a brane scenario described by two real scalar fields coupled with gravity, the second field contributing to propitiate internal arrangement, implementing therefore the so-called Bloch brane model. This framework was then explored by several authors in different situations, such as in the study of criticality and degeneracy, fermion localization, gauge field localization, graviton resonances, and so forth \cite{Dutra2005,Dutra,Almeida,Castro,Correa,FL1,FL2,Xie,Cruz,Cruz2014,Zhao}. 

Recently, it was suggested a new context of braneworld characterized by a hybrid behavior \cite{FC,AHB,CE}, which occurs when the scalar source supports localized structure with compactlike profile. In this scenario, it is manifested a thin brane behavior when the extra dimension is outside a compact domain, while a thick brane is revealed inside the compact region. In particular, the Ref. \cite{FC} proposed a route to smoothly go from kinks to compactons, leading to a braneworld formulation generated by compact-like defects; it occasioned a hybrid brane symmetric configuration. Phenomenological implications of this scenario are discussed in \cite{CE}, and in Ref. \cite{AHB} these ideas are extended to consider models that support asymmetric compactlike structures, in way that arises an asymmetric hybrid brane profile. 

In the current study we deal with the case of a flat brane, that is, with a brane with Minkowski internal geometry, but the subject is related to several other issues, in particular with investigations of bent branes, engendering an anti de Sitter or de Sitter internal geometry \cite{FN,dsads}, with the domain-wall/brane-cosmology correspondence \cite{A,B,C} and with the new concept of holographic cosmology \cite{T,HC} which is motivated by the AdS/CFT correspondence.

Inspired by these premises, in this work we are interested to investigate models described by potentials that support hybrid brane solutions developing internal structures, and this is what we call the hybrid Bloch brane. To implement this possibility, in Sec.~\ref{sec-1} we include necessary conditions to obtain Bloch branes in warped geometry. In Sec.~\ref{sec-2} we propose a route to get a hybrid Bloch brane from a thick Bloch brane. Firstly, we contemplate the case of a symmetric configuration, which gives us analytical results in the compact limit. We then move on and treat numerically an asymmetric case. 

The models studied in Sec.~\ref{sec-2} offer explicit possibilities to implement the hybrid Bloch brane scenario, and the results open up new issues, such as the ones investigated before in the case of the Bloch brane, namely the study of degeneracy and criticality, fermion localization, gauge field localization, and graviton resonances \cite{Dutra2005,Dutra,Almeida,Castro,Correa,FL1,FL2,Xie,Cruz,Cruz2014,Zhao}.  
We end the work in Sec. \ref{sec-com}, with conclusions and discussions.

\section{Bloch brane formulation}
\label{sec-1}

We now incorporate scalar fields into a warped geometry with one single extra dimension described by the line element
\be
ds^2=g_{ab}dx^a dx^b=\e^{2A}\eta_{\mu\nu}dx^{\mu}dx^{\nu}-dy^2,
\ee
where $a,b=0,1,2,3,4$; $y$ is the extra dimension, $A = A(y)$ is the warp function,  and $\eta_{\mu\nu}$ describes the  four-dimensional Minkowski spacetime $(\mu, \nu = 0, 1, 2, 3)$. In this case, the action is written as
\ben
I&=&\int{d^4xdy\sqrt{|g|}}\left(-\frac14 {R}+{\cal L}_{s}\right), 
\een
where ${\cal L}_s$ is used to describe the source scalar fields, that is,
\be
{\cal L}_{s}=\frac{1}{2}{\partial_a\phi\partial^a\phi}+\frac{1}{2}{\partial_a\chi\partial^a\chi}-V(\phi,\chi).
\ee
As usual, we assume that the scalar fields only depend on the extra dimension $y$. Under this circumstance, the equations of motion are expressed as
\bens
\phi''+4A'\phi'&=&\frac{\partial V}{\partial \phi}, \\
\chi''+4A'\chi'&=&\frac{\partial V}{\partial \chi}, \\
A''&=&-\frac{2}{3}\left(\phi'^2+\chi'^2\right), \\
A'^2&=&-\frac{1}{6}\left(\phi'^2+\chi'^2\right)-\frac{1}{3}V,
\eens
where prime denotes differentiation with respect to the coordinate $y$. If we focus on the first-order framework, with interest to obtain first-order differential equations, we write the potential in terms of a superpotential $W=W(\phi,\chi)$, such that
\be
\label{pot5}
V(\phi,\chi)=\frac{1}{2}W_{\phi}^2+\frac{1}{2}W_{\chi}^2-\frac{4}{3}W^2,
\ee
with $W_{\phi}=\partial W/\partial \phi$, and $W_{\chi}=\partial W/\partial \chi$. In this case, the equations of motion are reduced to the following first-order equations
\ben \label{eqq}
\phi' = W_{\phi}, \:\:\:\:\: \chi' = W_{\chi}, \:\:\:\:\: \mbox{and} \:\:\:\:\: A' = -\frac{2}{3}W.
\een
Moreover, the energy can be written in terms of the energy density in the form
\be  
E=\int_{-\infty}^{\infty}\rho(y)\, dy,
\ee
where
\be
\rho(y)=\e^{2A}\left(W_{\phi}^2+W_{\chi}^2-\frac{4}{3}W^2\right).
\ee
One notes that for field configurations that solve the first-order equations, it is possible to write the energy density as
\be\label{eneden}
\rho(y)=\frac{d}{dy} (e^{2A}W),
\ee 
and so the total energy vanishes, because asymptotically the warp factor $e^{2A}$ suppress the superpotential $W(\phi,\chi)$, for the physically acceptable solutions of the first-order equations.  

\subsubsection{Linear stability}

In the braneworld context, the general treatment to study stability of the gravity sector is done through small perturbations of the metric and the scalar fields. The perturbed metric is
\be
ds^2=\e^{2A}\left(\eta_{\mu\nu}+\epsilon h_{\mu\nu}\right)dx^{\mu}dx^{\nu}-dy^2, 
\ee
and the scalars fluctuations are $\phi\rightarrow \phi+\epsilon \tilde{\phi}$ and $\chi\rightarrow \chi+\epsilon \tilde{\chi}$. Here $h_{\mu\nu} = h_{\mu\nu}(x^\mu, y)$, $\tilde{\phi}=\tilde{\phi}(x^\mu, y)$, and $\tilde{\chi}=\tilde{\chi}(x^\mu, y)$. 

Choosing the fluctuations of the metric as transverse and traceless \cite{FD}, $h_{\mu\nu}\rightarrow \bar{h}_{\mu\nu}$, the equation of motion for the fluctuations decouple from the source fields and satisfies a much simpler form
\be
\label{SP}
\left(\partial_y^2+4A'\partial_y-\e^{-2A}\Box\right)\bar{h}_{\mu\nu}=0,
\ee
where $\Box=\eta^{\mu\nu}\partial_\mu\partial_\nu$. The Eq.~\eqref{SP} can be recast in a Schr\"odinger-like equation through the change of coordinates $dz=\e^{-A(y)}dy$ and making $\bar{h}_{\mu\nu}(x, y)=\e^{ikx}\e^{-3A(z)/2}H_{\mu\nu}(z)$. Using these arguments, the expression for the metric fluctuations simplifies to
\be\label{SC}
\left(-\frac{d^2}{dz^2}+U(z)\right)H_{\mu\nu}(z)=k^2H_{\mu\nu}(z),
\ee
where we introduced the potential 
\be
U(z)=\frac{3}{2}\ddot{A}(z)+\frac{9}{4}\dot{A}^2(z)=\frac{3}{4}\e^{2A}\left(2A''+5A'^{2}\right).
\ee
Here the dot is used to represent derivative with respect to the coordinate $z$. With this one writes
\be  
-\frac{d^2}{dz^2}+\frac32 {\ddot A}+\frac 94{\dot A}^2=\left(-\frac{d}{dz}-\frac32{\dot A}\right)\left(\frac{d}{dz}-\frac32{\dot A}\right),
\ee
and this shows that the above Eq.~\eqref{SC} cannot support negative bound states; thus, $k^2\geq 0$ and $k$ is real, rendering the braneworld scenario stable against fluctuations in the metric.

\section{From thick Bloch brane to hybrid Bloch brane}\label{sec-2}

Let us now focus on the hybrid Bloch brane scenario. We first recall that the Bloch brane scenario suggested in Ref.~\cite{BB} is described under the first-order framework, with the choice
\be  
W(\phi,\chi)=\phi -\frac13 \phi^3-r \phi\chi^2
\ee
where $r$ is a real parameter that can vary in the interval $r\in(0,1)$. In the investigations that follows we introduce two distinct possibilities that modify the Bloch brane model, leading us to two distinct scenarios that follow the hybrid profile suggested before in \cite{FC,AHB}. 

\subsection{Model 1}

\begin{figure}[t]
\includegraphics[width=4.2cm,height=4.2cm]{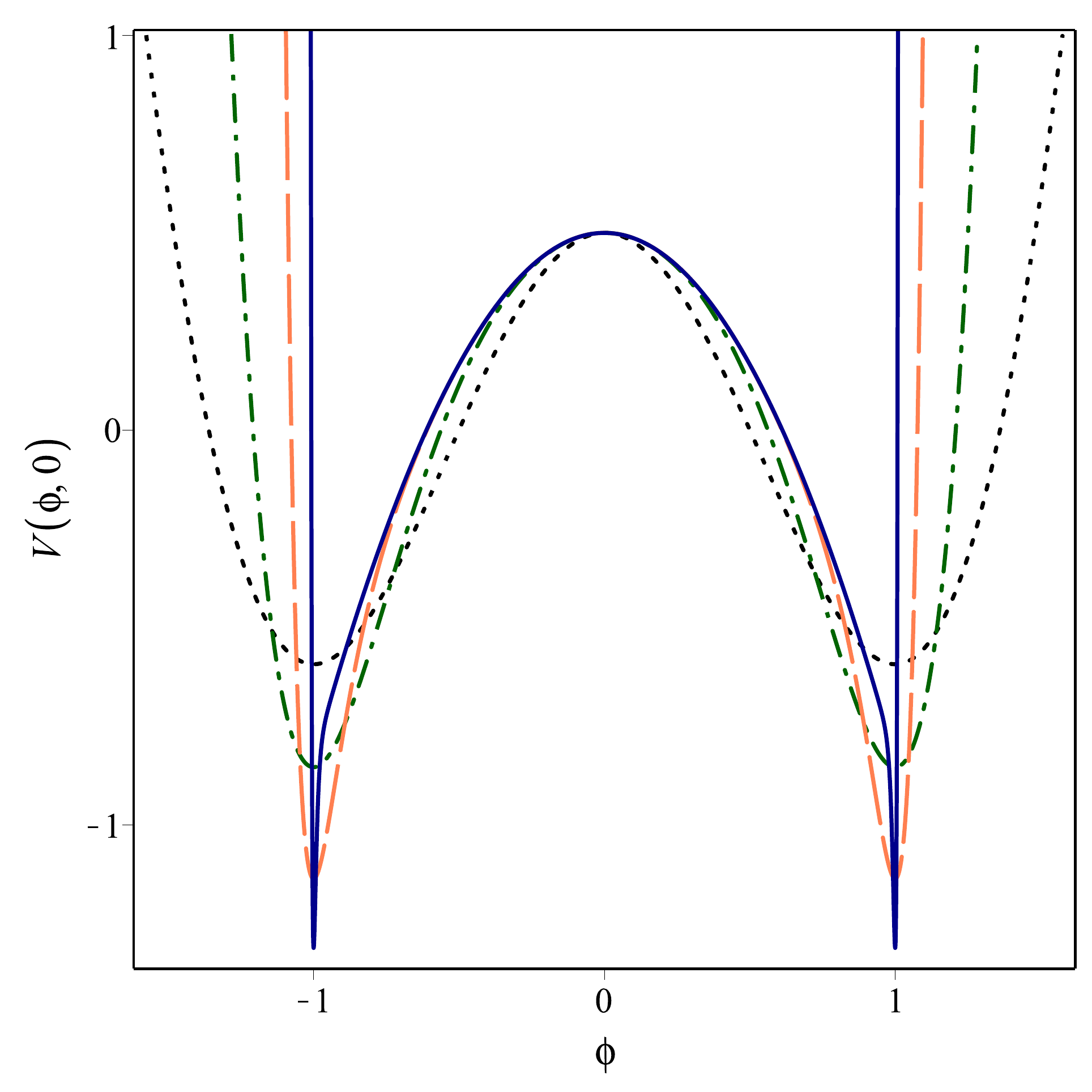}
\includegraphics[width=4.2cm,height=4.2cm]{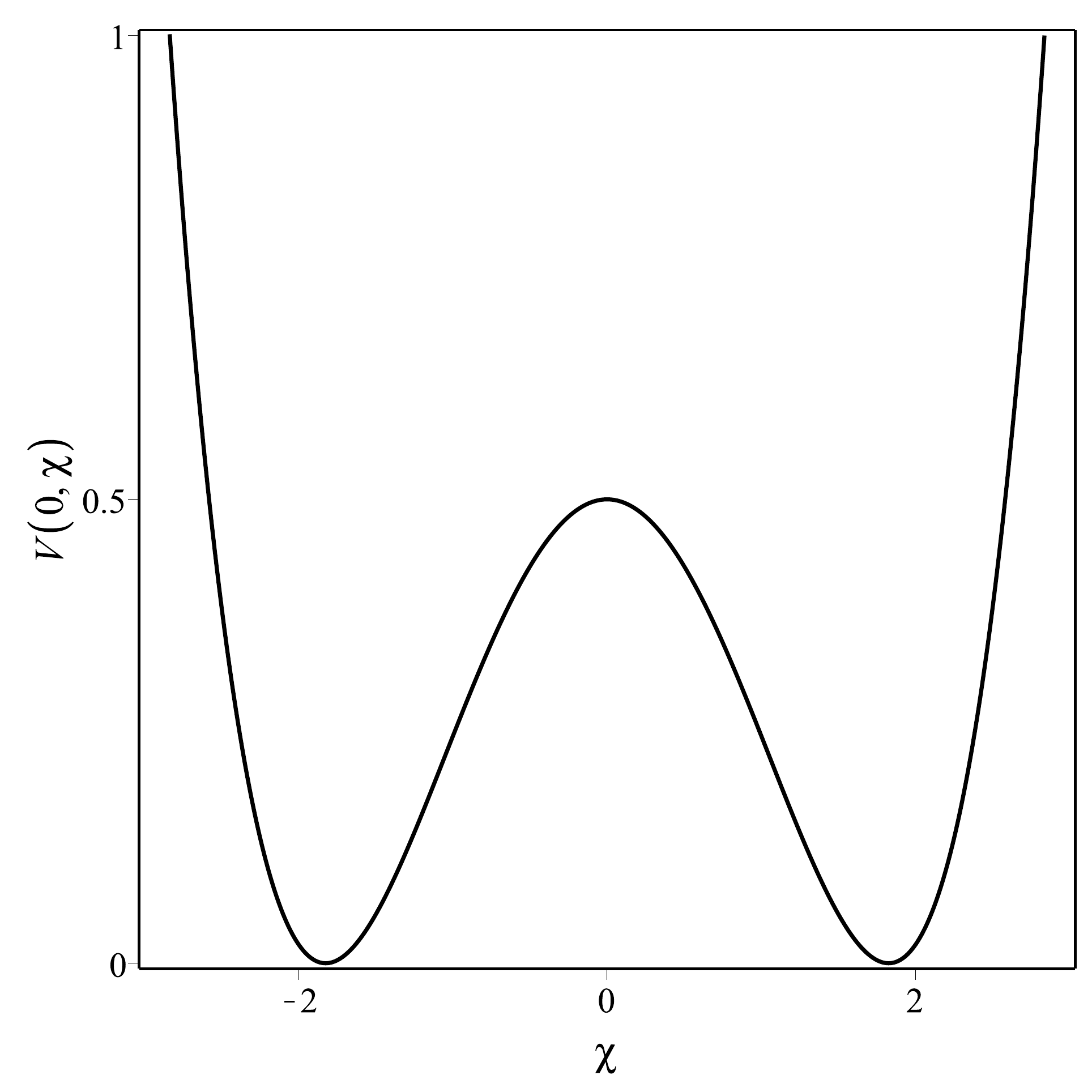}
\caption{In the left panel, we display the potential \eqref{pp0} versus $\phi$, for $n=1,2,4,60$ depicted with dotted (black), dot-dashed (green), dashed (red), and solid (blue) lines, respectively. In the right panel we show the potential \eqref{pq0} versus $\chi$, for $r=0.3$.}
\label{fig:1}
\end{figure}

We start with the symmetric model described by the following function $W$ 
\be
W(\phi,\chi)=\phi-\frac{\phi^{2n+1}}{2n+1}-r\phi\chi^2, 
\ee
with $n$ being a positive integer; for $n = 1$, we get back to the model \cite{BNRT} which reproduces the thick Bloch brane model \cite{BB}. According to Eq.~\eqref{pot5}, the brane potential presents the following profiles:
\bens \slabel{pp0}
V(\phi,0)&=&\frac{1}{2} (1-\phi^{2n})^2-\frac{4}{3}\phi^2\left(1-\frac{\phi^{2n}}{2n+1}\right)^2\\
\slabel{pq0}
V(0,\chi)&=&\frac{1}{2}(1-r\chi^{2})^2.
\eens
In this case, we see that both $V(\phi,0)$ and $V(0,\chi)$ present spontaneous symmetry breaking, as displayed in Fig. \ref{fig:1}, and this is consistent with appearance of internal structure inside defects in the above system.  Additionally, the vertical cross section of the potential at  $\chi=0$ possesses a behavior that induces the presence of compact structures, with $V(\pm1,0)=-16n^2/3(2n+1)^2$, which becomes $-4/3$ for large values of $n$.

\begin{figure}[t]
\includegraphics[width=4.2cm,height=4.2cm]{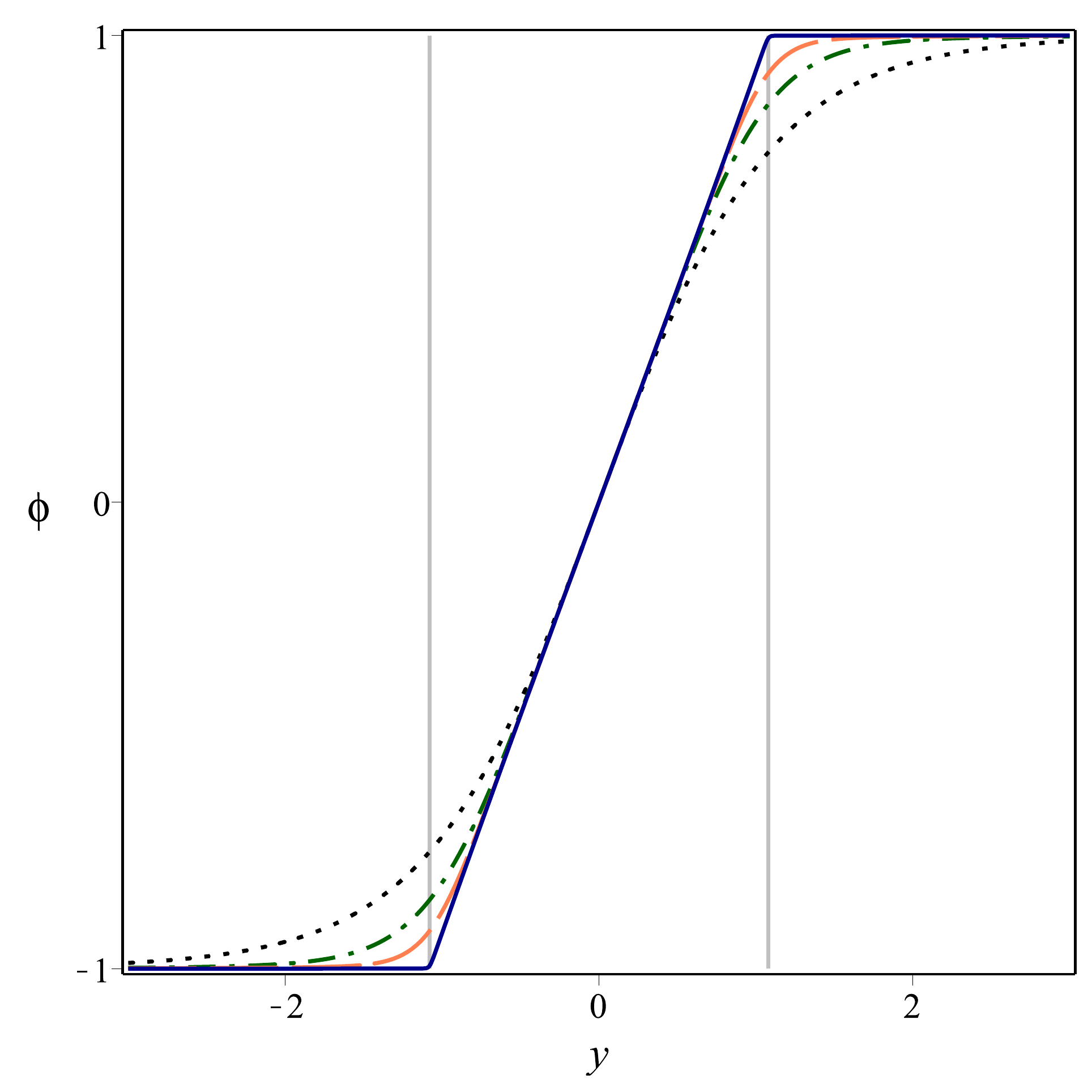}
\includegraphics[width=4.2cm,height=4.2cm]{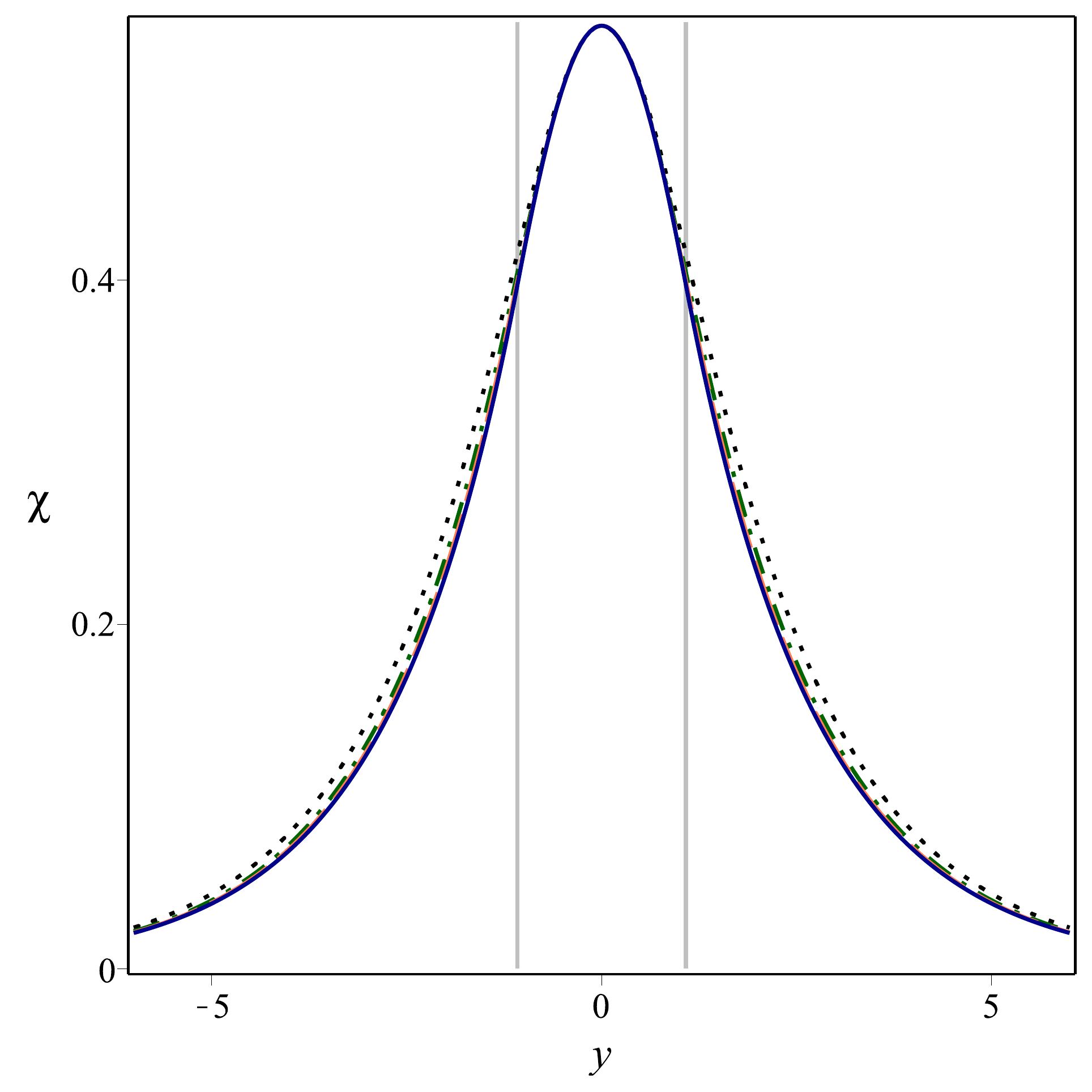}
\includegraphics[width=4.2cm,height=4.2cm]{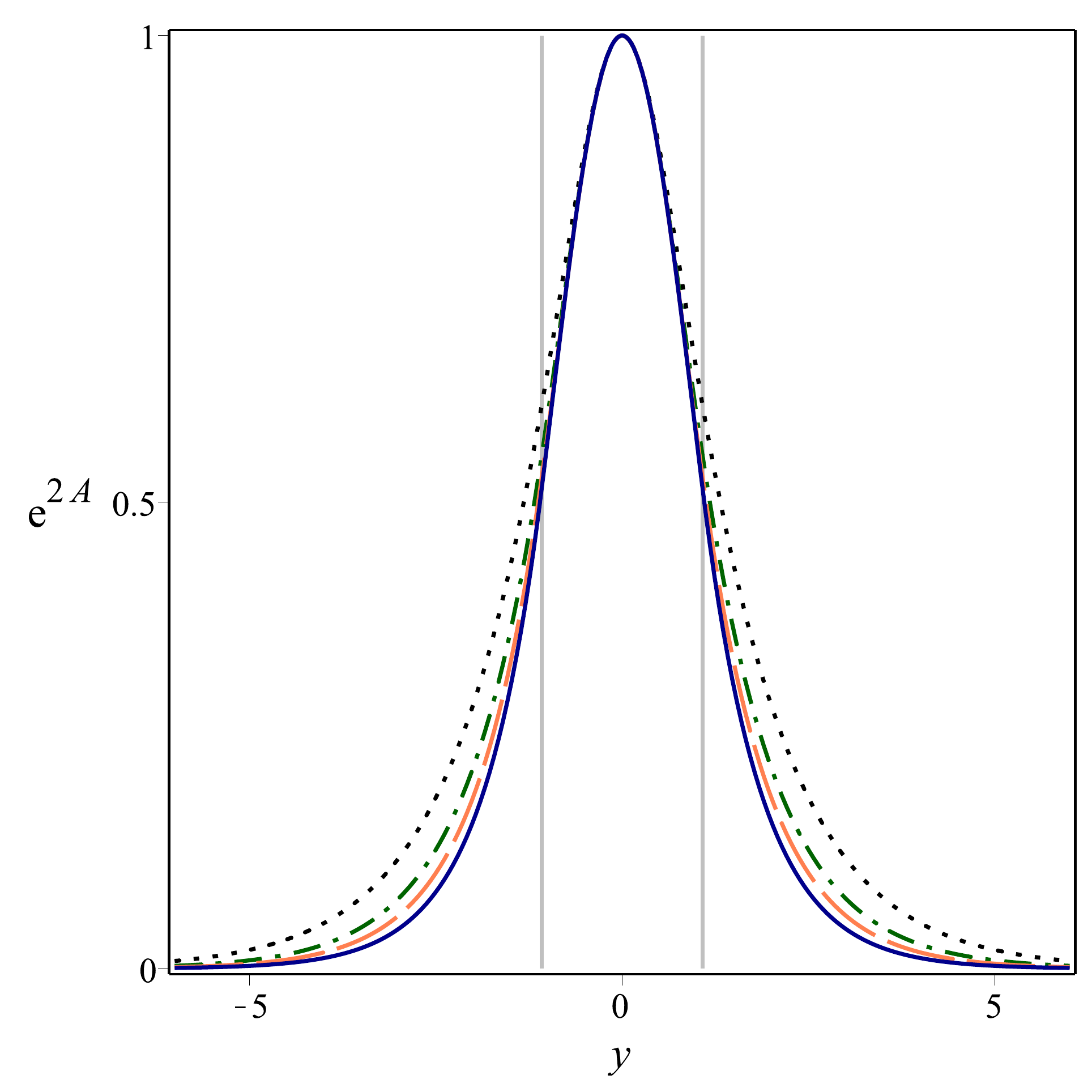}
\includegraphics[width=4.2cm,height=4.2cm]{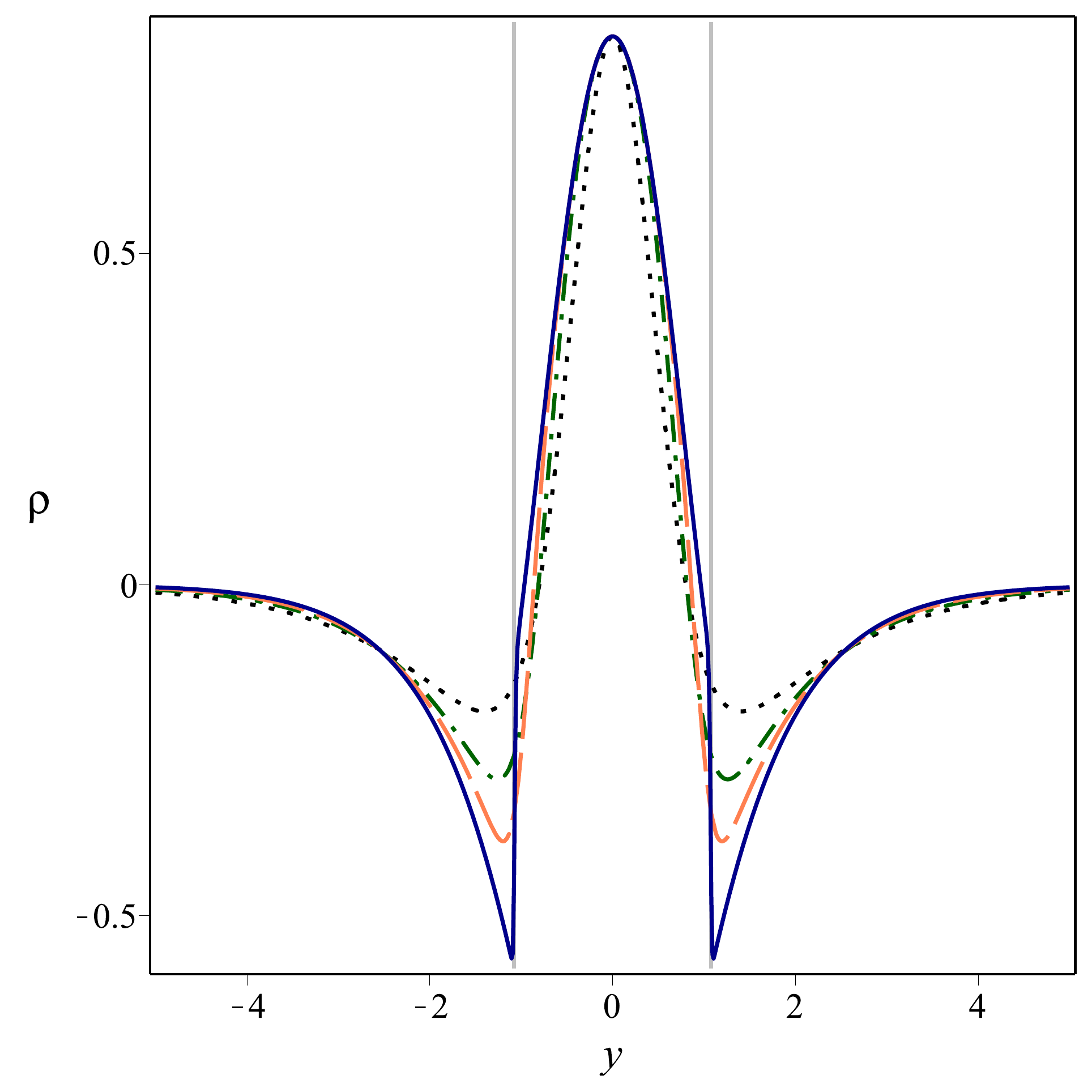}
\caption{In the top panel, we display the fields solutions $\phi(y)$ (left) and $\chi(y)$ (right). In the bottom panel we show the warp factor
$\exp[2A(y)]$ (left) and energy density $\rho(y)$ (right). We consider the same values of $n$ used in Fig. \ref{fig:1}, and take $r=0.3$.}
\label{fig:2}
\end{figure}

\begin{figure}[t]
\includegraphics[width=4.2cm,height=4.2cm]{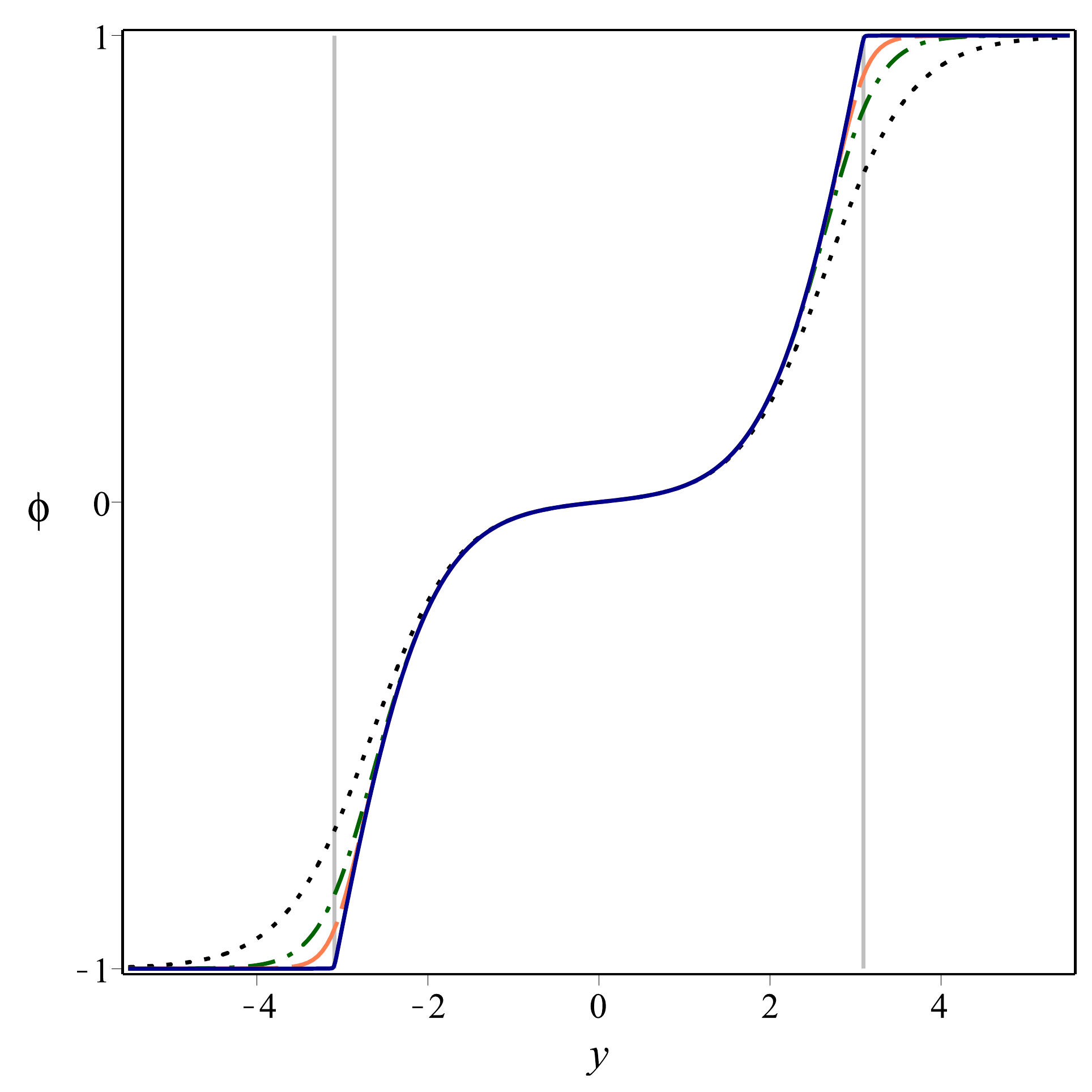}
\includegraphics[width=4.2cm,height=4.2cm]{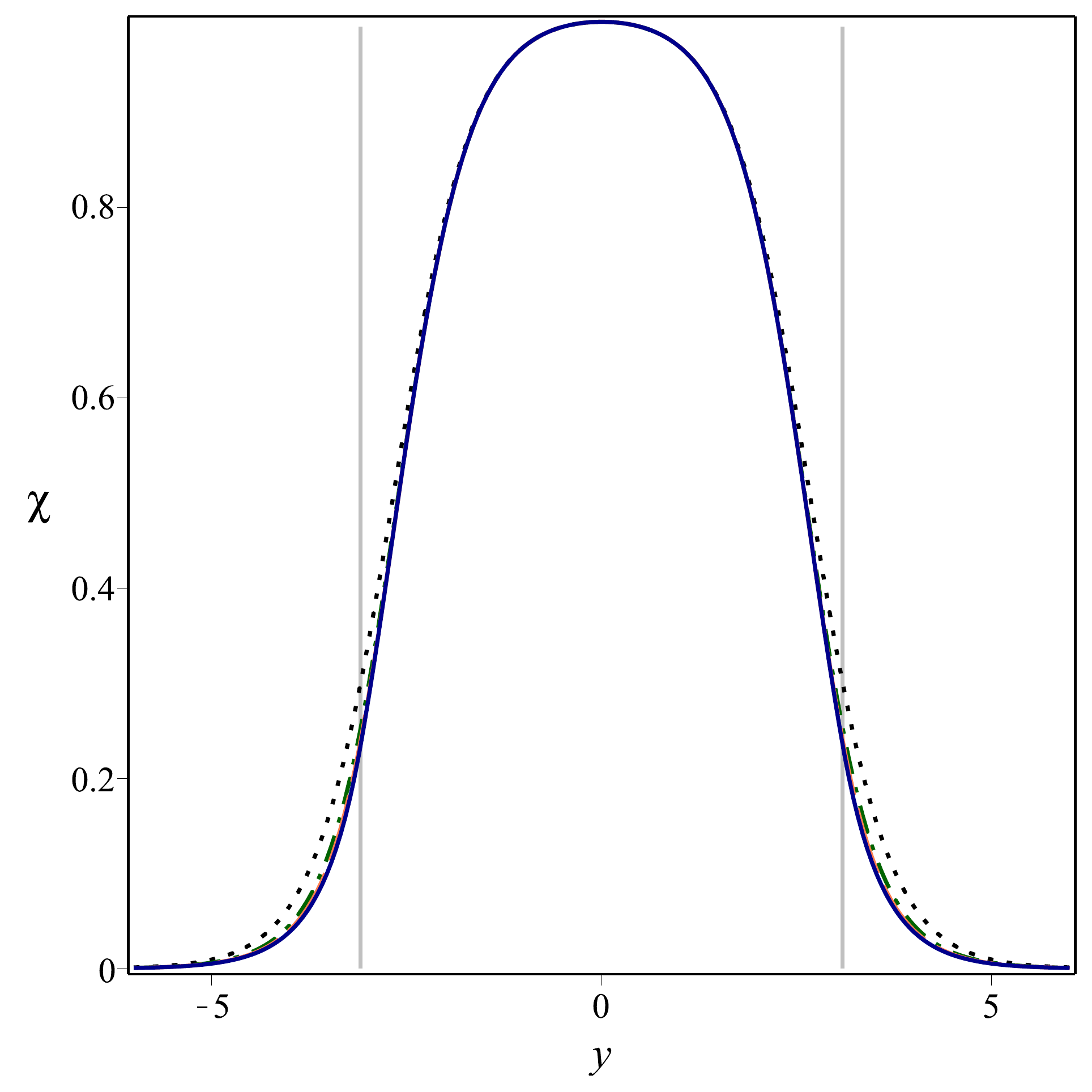}
\includegraphics[width=4.2cm,height=4.2cm]{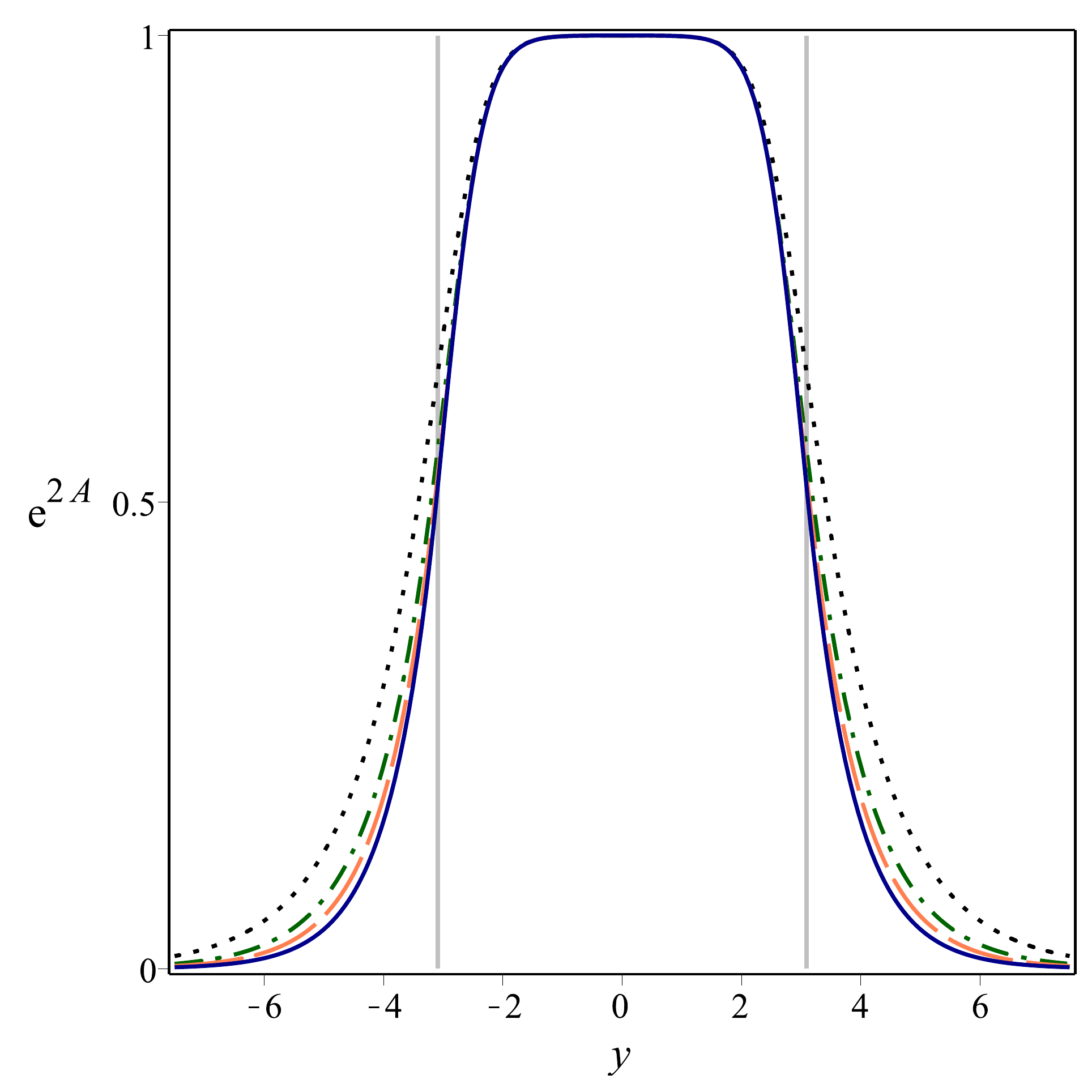}
\includegraphics[width=4.2cm,height=4.2cm]{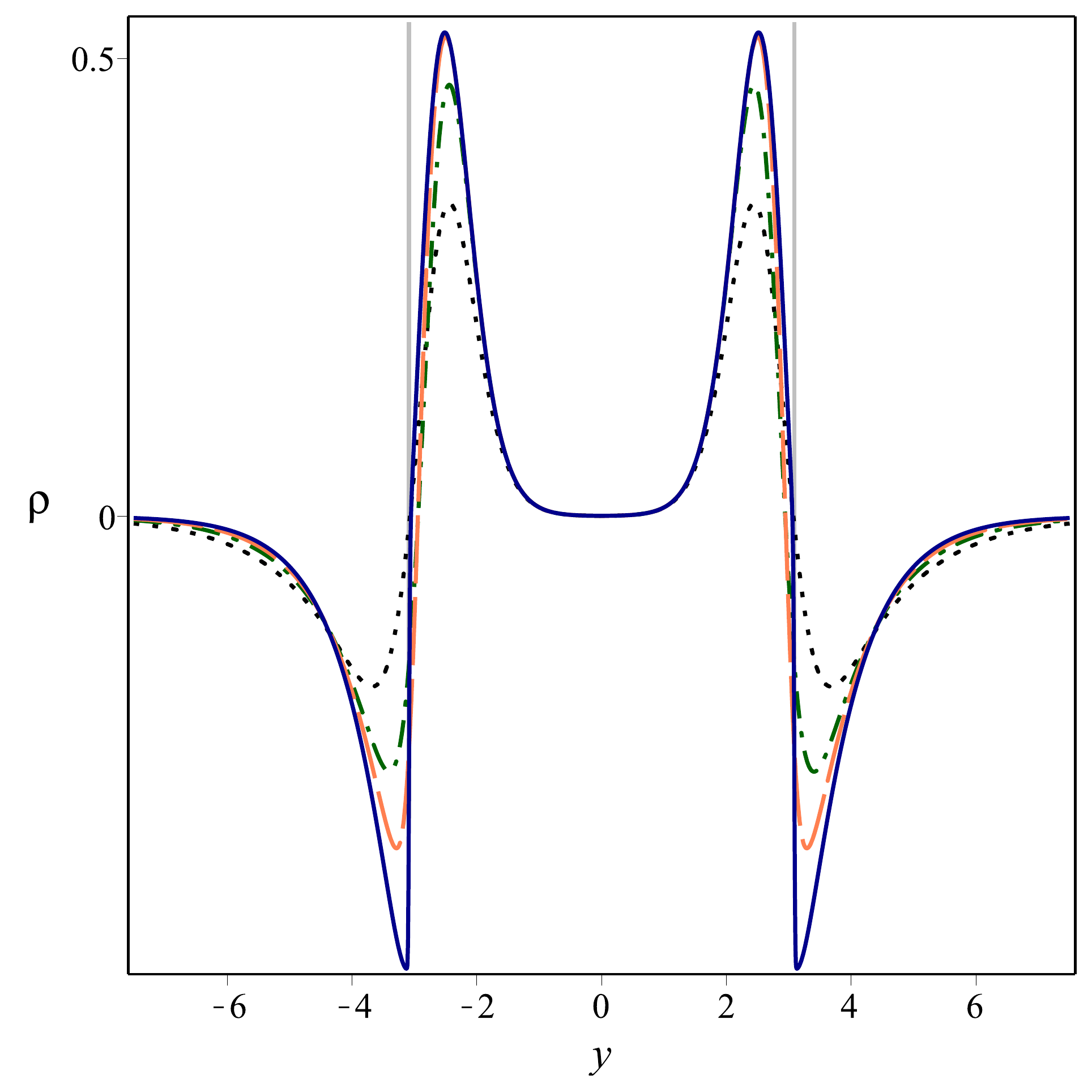}
\caption{The same solutions, warp factor and energy density of Fig.~\ref{fig:2}, displayed for the same values of $n$, but now with $r=0.99$.}
\label{fig:3}
\end{figure}

The solutions of the first-order Eqs.~\eqref{eqq} together with their respective energy densities are shown in Figs.~\ref{fig:2} and \ref{fig:3}, for various values of $n$, keeping fixed the parameter $r$ which controls interactions between the two scalar fields. These figures show kinklike solutions for the field $\phi(x)$ and lumplike solutions for $\chi(x)$, at the top left and right panels, respectively. They also show the warp factor and the energy density at the bottom left and right panels, respectively. In the limit where $n$ is very large, the kinklike solution tends to become compacted to a finite region of space, into a compact interval, while the lumplike solution remains decaying asymptotically. When the extra dimension is outside the compact interval, arises a thin brane behavior, identified by the non-trivial behavior of the energy density, while a thick brane is displayed inside the compact space \cite{FC,AHB,CE}. And this is the reason to suggest that the compact-like behavior leads to the hybrid brane, as it was originally suggested in \cite{FC} in the case of a single field. Here, however, one is changing the Bloch brane \cite{BB} into a hybrid Bloch brane.

In particular, we can see in Fig. \ref{fig:3} that for $r$ very close to unit, for small values  of $n$ one is led to the emergence of the 2-kink profile for the $\phi$ field, and this enlarges the thickness of $\chi$ and of the warp factor, splitting the energy density. These are interesting features, which appear in systems supporting thick branes solutions that develop internal structure \cite{TK,BF,BB,Dutra}, as well as in systems which deal with critical phenomena of thick branes generated at high temperature \cite{AC}. Furthermore, as $n$ increases to larger and larger values, the 2-kink profile gives rise to new structures, of the form of two half-compact solutions, as it was first identified in Ref. \cite{AHB}. We recall that a half-compact solution behaves as a compact solution at one side, and as a standard solution at the other side. 

In order to better understand the hybrid Bloch brane profile, we take the limit $n \rightarrow \infty$. It gives
\be
W(\phi,\chi) =\phi-r\phi\chi^2.
\ee
Considering first the case $|y|\leq \bar{y}$, where we designate $\bar{y}$ as the point where $\phi$ reaches the value $1$, the first-order equations \eqref{eqq} take the form
\ben
\phi'&=&1-r\chi^2, \:\:\:\: \chi'=-2r\phi\chi, \:\:\mbox{and} \:\:\:\: A'=-\frac{2}{3}\phi(1-r\chi^2). \nonumber \\
\een
Eliminating $y$ from the two first equations above we have
\be
\frac{d\phi}{d\chi}=-\frac{1-r\chi^2}{2r\phi\chi}.
\ee
The solution of this equation provides a trajectory on $(\phi, \chi)$ plane,
\ben
\label{orbt}
\phi^2=c+\frac{1}{2}\chi^2-\frac{1}{r}\ln(\chi),
\een
where $c$ is an integration constant to be determined. We know that the component $\chi$ is maximum when $\phi$ goes toward zero, because of that we impose $\phi(y=0)=0$ and $\chi(y=0)=\sqrt{r}$. Then we find,
\be
\phi^2=\frac{1}{2r}\ln(\frac{r}{\chi^2})-\frac{1}{2}(r-\chi^2).
\ee
Since we are taking $0<r<1$, we can ensure real fields along the whole trajectory. Having in mind these premises, we get
\bens
\phi'&=&1+W_0\left(-r^2\e^{-2r\phi^2-r^2}\right),\\
\chi&=&\sqrt{-\frac{1}{r}W_0\left(-r^2\e^{-2r\phi^2-r^2}\right)},\\
A&=&-\frac{1}{3}\phi^2,
\eens
where $W_0$ is the principal branch of the Lambert function and $-r^2\e^{-2r\phi^2-r^2}\in (-1/e,0)$. The solution of $\phi$ may be expressed as a transcendental function,
\be
\phi+\frac{\sqrt{2\pi}}{4}\sum^{\infty}_{k=1}{\frac{k^{k-1/2}}{k!}r^{2k-1/2}\e^{-kr^2}\erf(\sqrt{2kr}\phi)}=y.
\ee
It implies that 
\be
\bar{y}=1+\frac{\sqrt{2\pi}}{4}\sum^{\infty}_{k=1}{\frac{k^{k-1/2}}{k!}r^{2k-1/2}\e^{-kr^2}\erf(\sqrt{2kr})}.
\ee

Now, we will analyse the situation $|y|> \bar{y}$. The first-order equations are
\be
\phi'=0, \:\:\:\: \chi'=-2r\frac{\phi}{|\phi|}\chi, \:\:\mbox{and} \:\:\:\: A'=-\frac{2}{3}\frac{\phi}{|\phi|}(1-r\chi^2). \nonumber \\
\ee
Using appropriate boundary conditions, the solutions for these equations are
\bens
\phi(y)&=&\frac{y}{|y|}, \\
\chi(y)&=&\sqrt{-\frac{1}{r}W_0\left(-r^2\e^{-2r-r^2}\right)}\e^{-2r(|y|-\bar{y})}, \\
A(y)&=&-\frac{2}{3}(|y|-\bar{y})-\frac{1}{3}+\frac{1}{6r}W_0\left(-r^2\e^{-2r-r^2}\right)\times \nonumber \\ &&\left[\e^{-4r(|y|-\bar{y})}-1\right].
\eens
In fact, from the last above expression we see the warp factor decaying as a thin brane when the extra dimension is outside the compact space $[-\bar{y},\bar{y}]$.

\begin{figure}
\includegraphics[width=4.2cm,height=4.2cm]{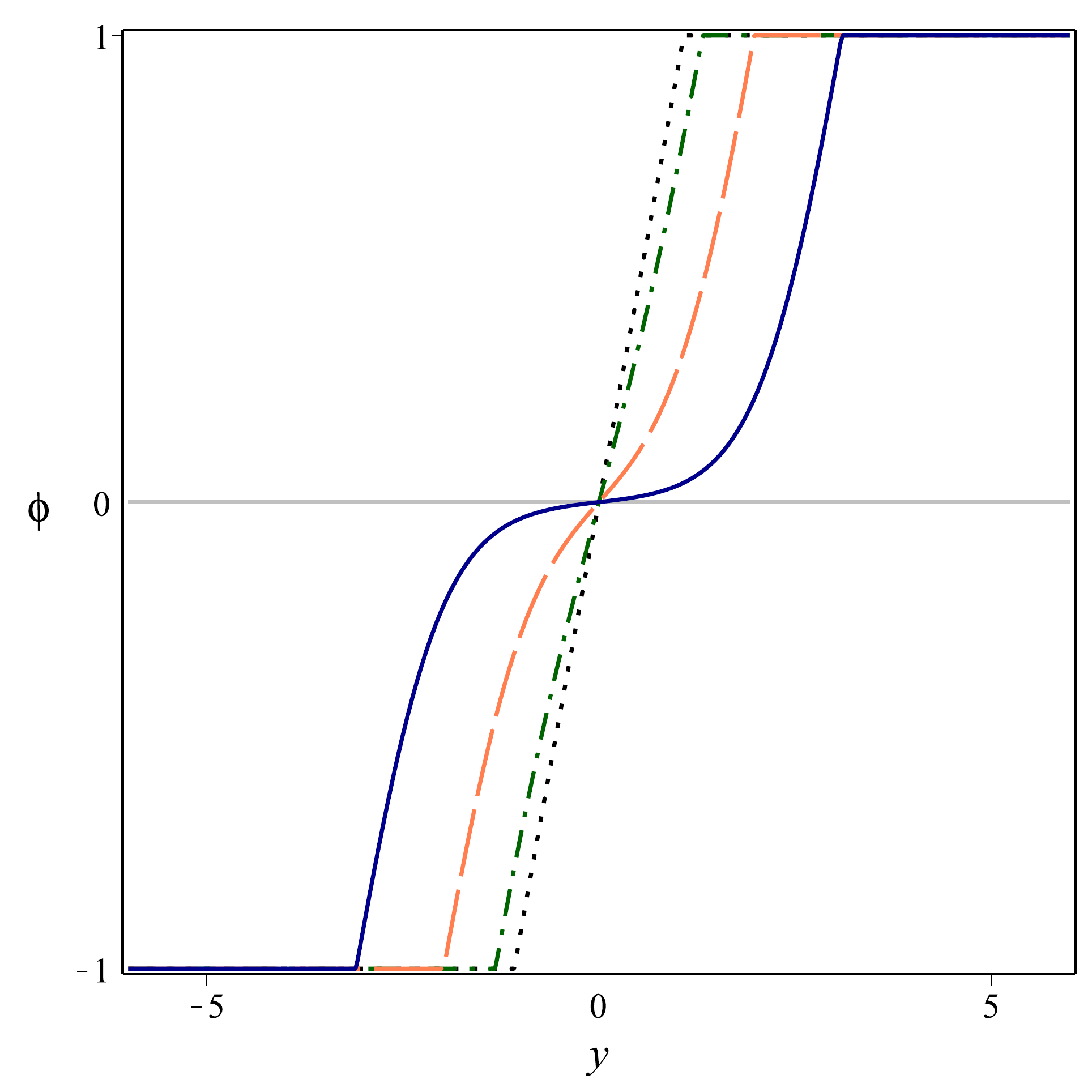}
\includegraphics[width=4.2cm,height=4.2cm]{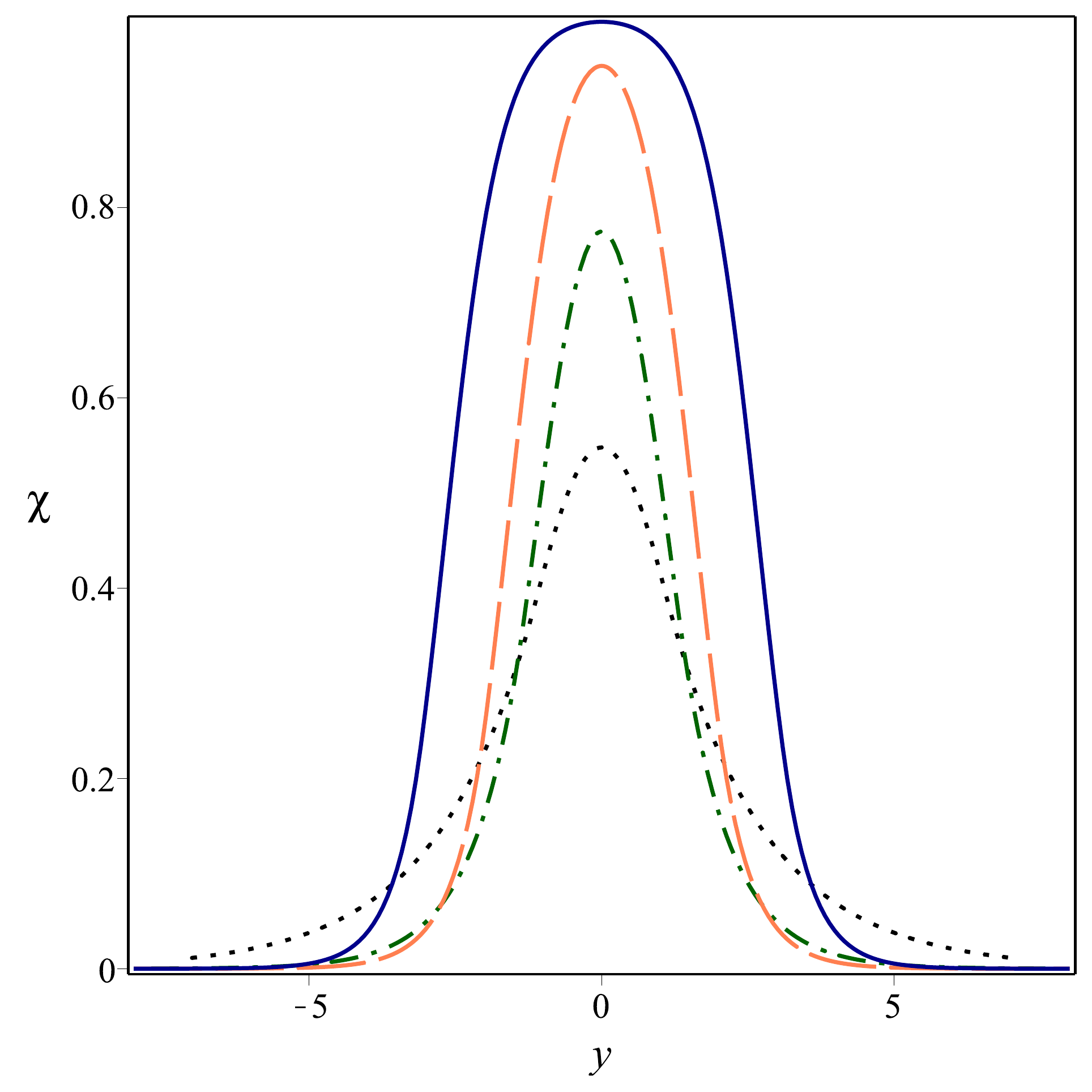}
\includegraphics[width=4.2cm,height=4.2cm]{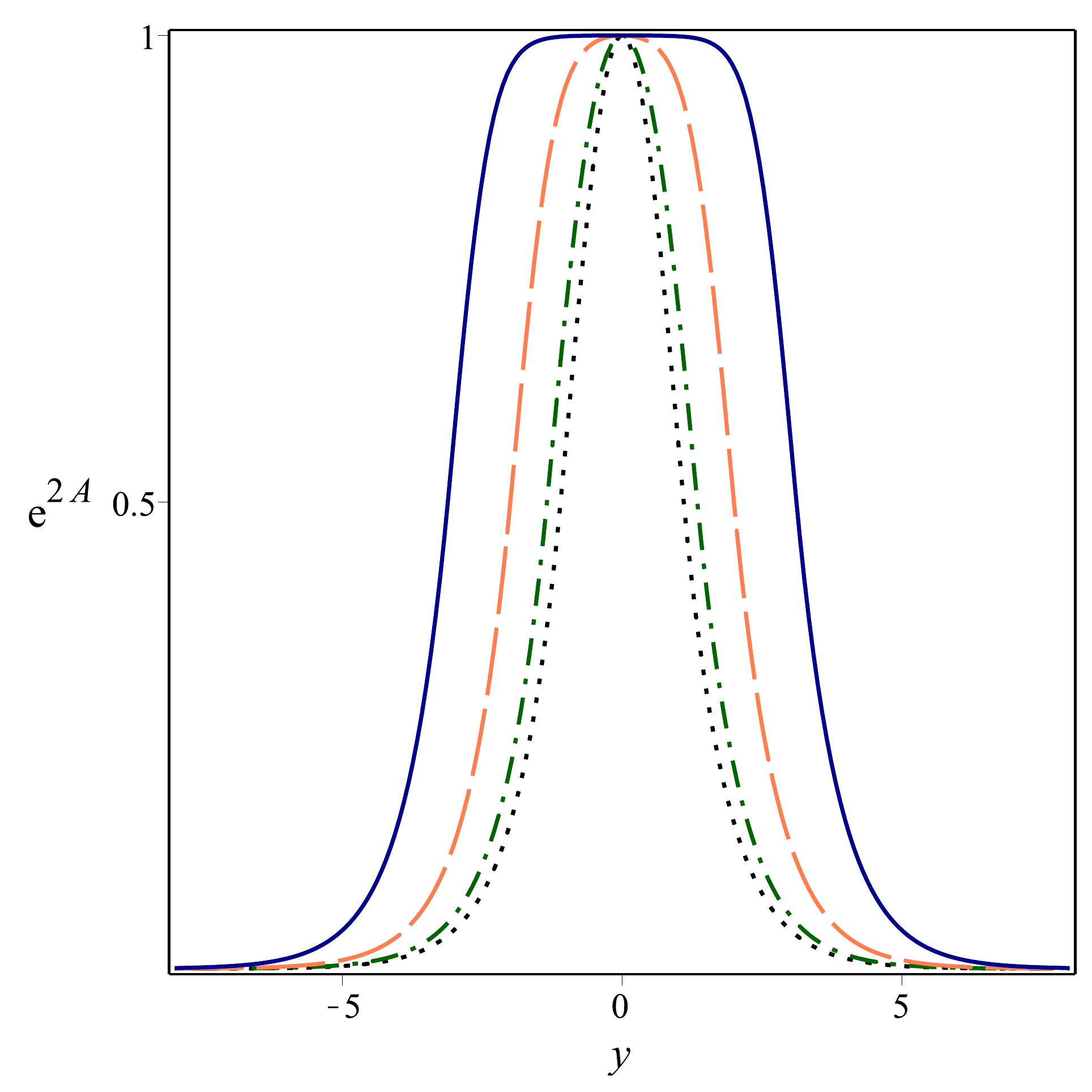}
\includegraphics[width=4.2cm,height=4.2cm]{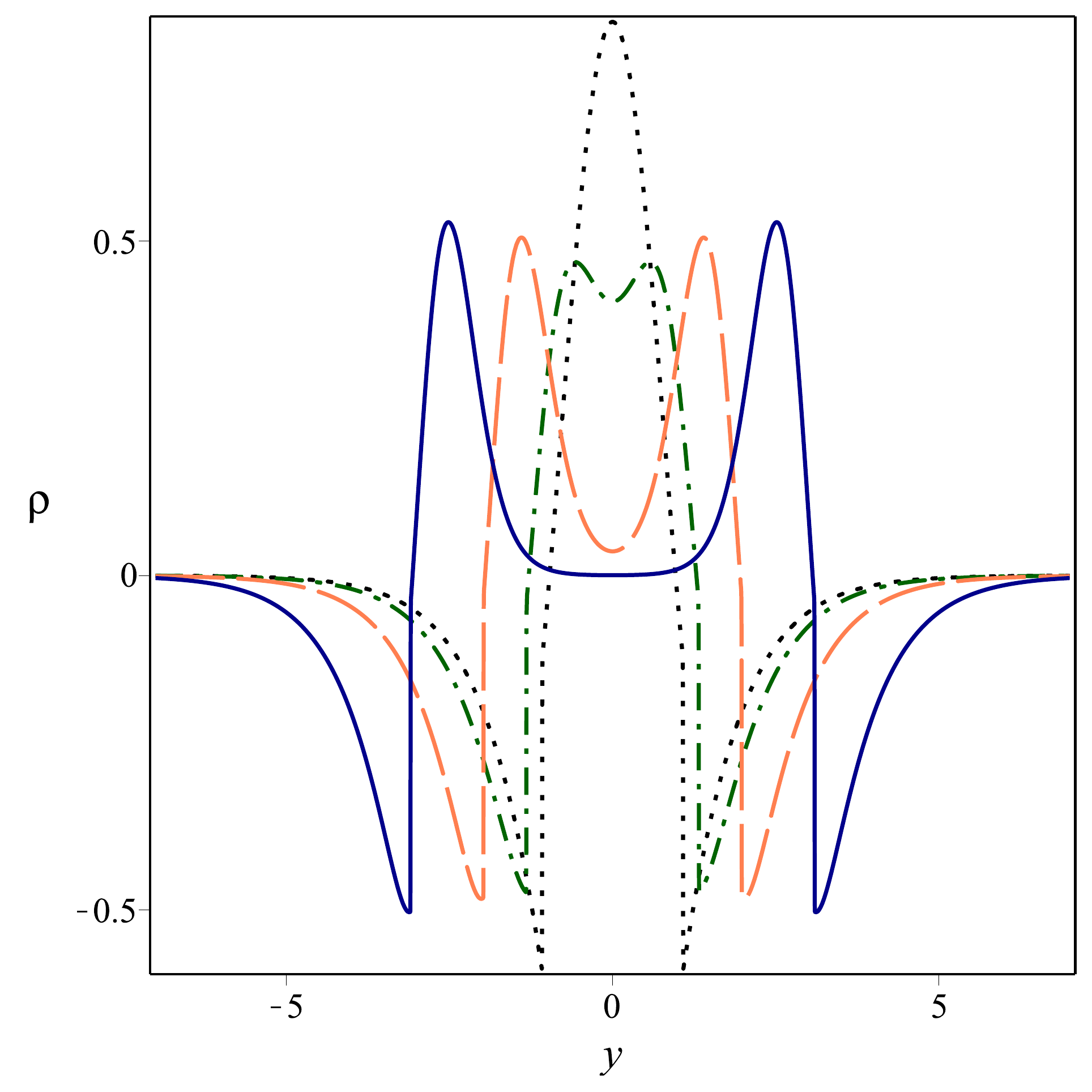}
\caption{The hybrid Bloch brane for $r=0.3, 0.6, 0.9, 0.99$, depicted with dotted (black), dot-dashed (green), dashed (red), and solid (blue) lines, respectively. In the top panel, the field solutions $\phi(y)$ and $\chi(y)$, and in the bottom panel, the warp factor
$\exp[2A(y)]$ and energy density $\rho(y)$.}\label{fig:4}
\end{figure}

In Fig. \ref{fig:4} we depict the analytical results obtained above, for some values of $r$. These results shown an internal structure being exhibited in the hybrid brane. It is observed an increase of brane thickness as $r$ goes to unity. Moreover, as $r$ increases, the compact kink behaves as two half-compact structures; at the same time occurs a change of energy density behavior where appears a splitting of its maximum into two new maxima, identifying two interfaces induced by the appearance of internal structure, due to the two half-compact behavior. 

The energy density of the hybrid Bloch brane has a discontinuity at $|y|=\bar{y}$, given by
\be
\rho(\bar{y}^-)-\rho(\bar{y}^+)=\e^{-2/3}\left[1+W_0\left(-r^2\e^{-2r-r^2}\right)\right]^2,
\ee
where $\bar{y}^-$ and $\bar{y}^+$ symbolize the limits when $y$ tends to $\bar{y}$ from left and right sides, respectively. However, it is integrable and develops zero total energy, once the energy density can be expressed as in Eq.~\eqref{eneden}.

\subsection{Model 2}

\begin{figure}
\includegraphics[width=4.2cm,height=4.2cm]{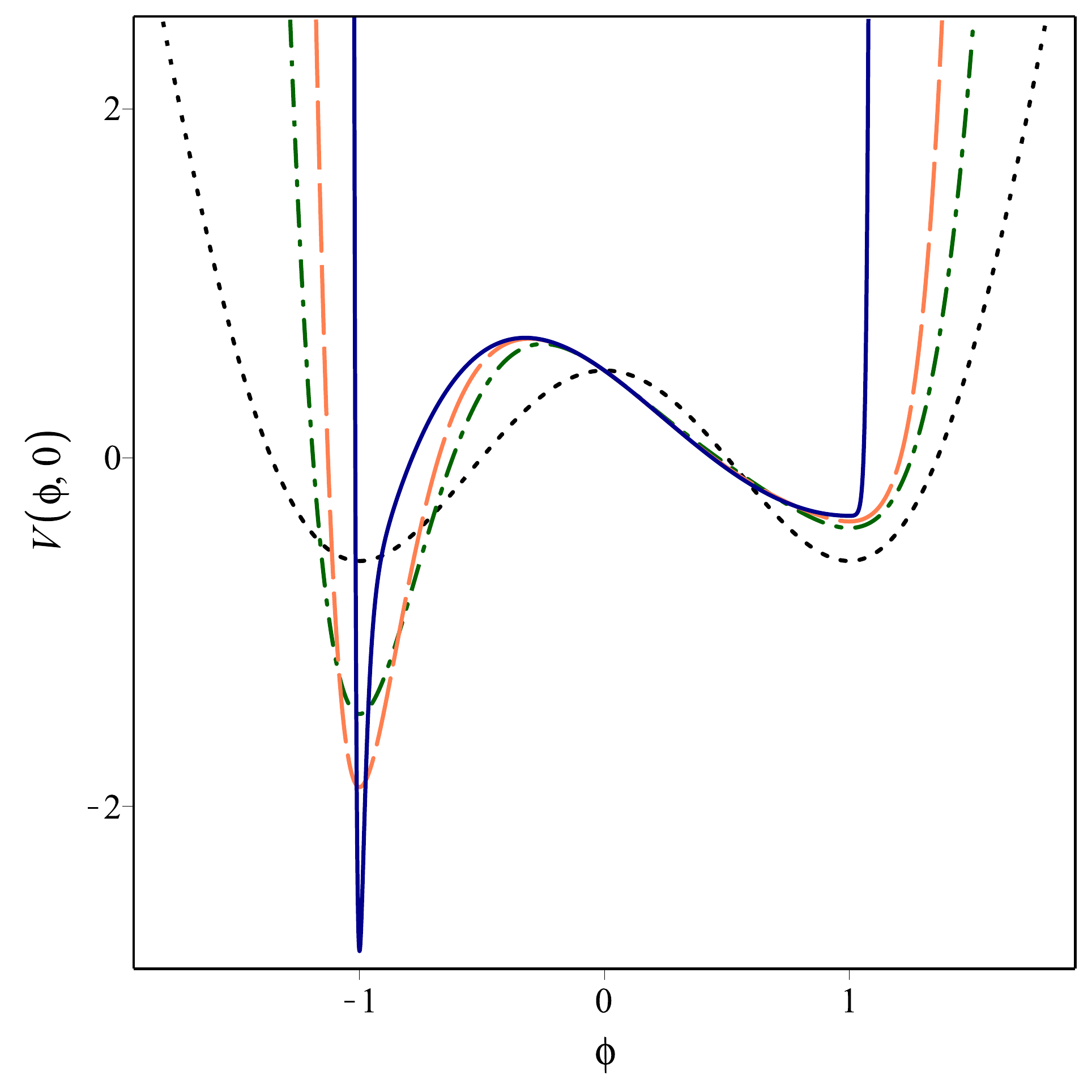}
\includegraphics[width=4.2cm,height=4.2cm]{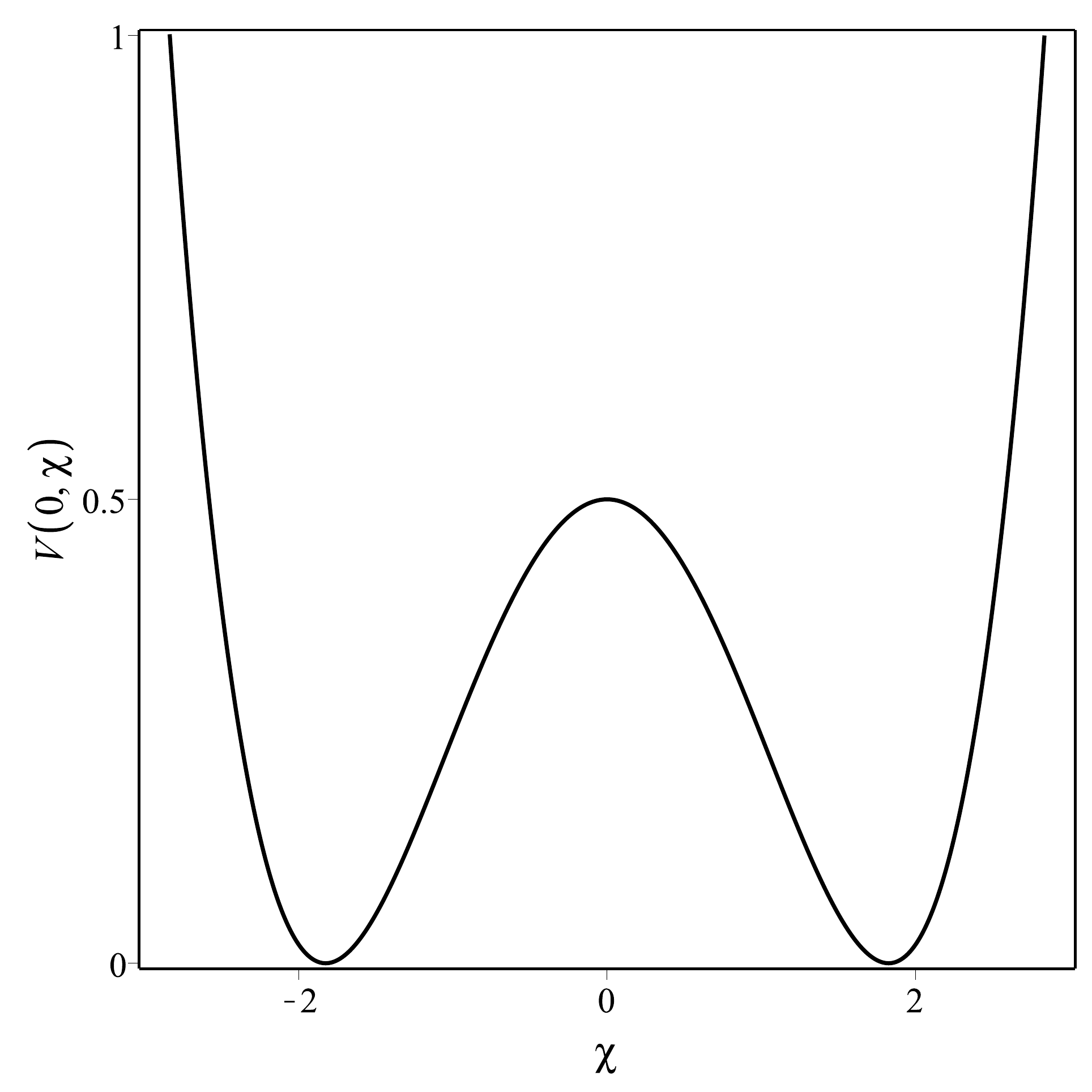}
\caption{In the left panel, we display the potential \eqref{pp01} versus $\phi$ for $p=1,3,5,45$, depicted with dotted (black), dot-dashed (green), dashed (red), and solid (blue) lines, respectively. In right panel, we show the potential \eqref{pq01} versus $\chi$, with $r=0.3$.}
\label{fig:5}
\end{figure}

\begin{figure}
\includegraphics[width=4.2cm,height=4.2cm]{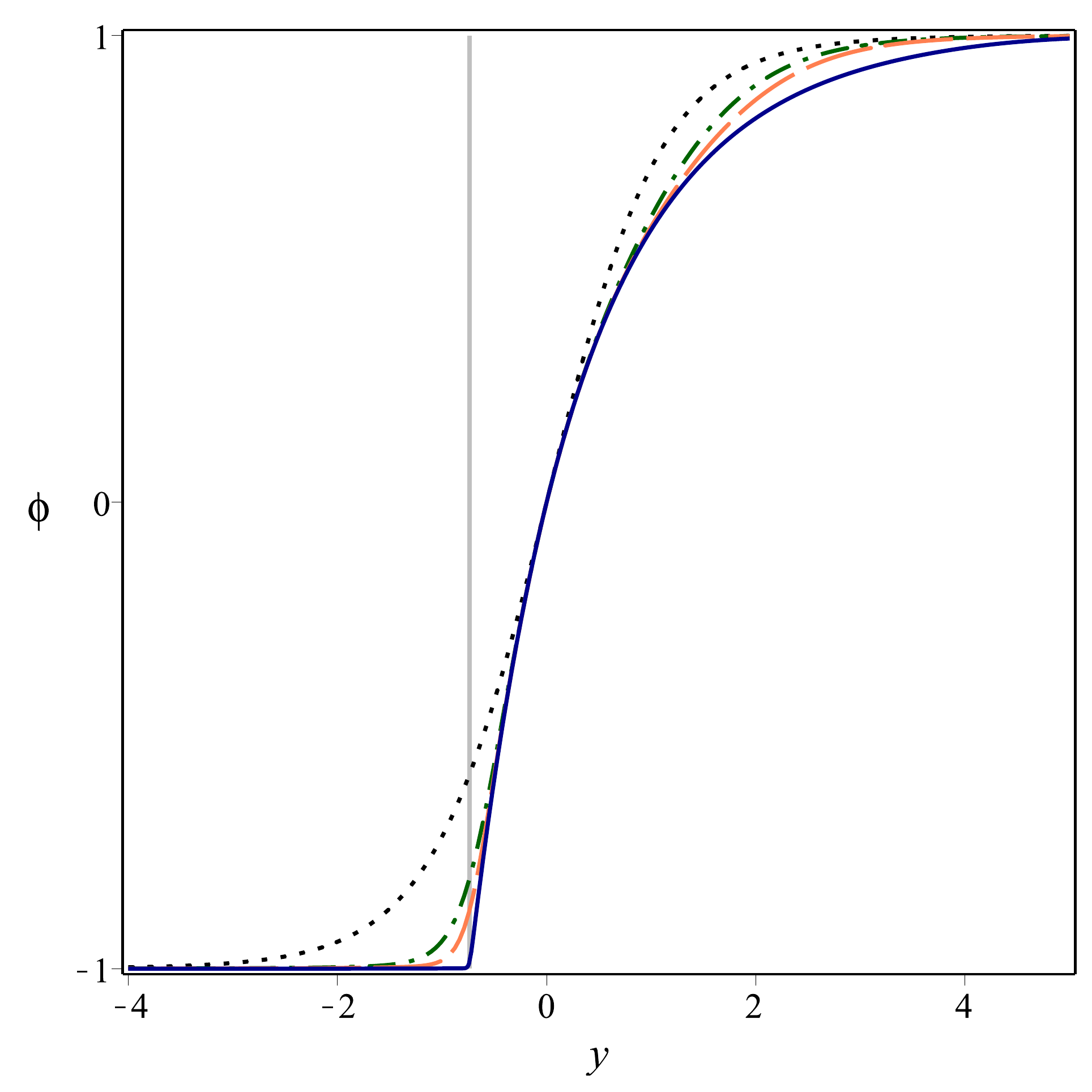}
\includegraphics[width=4.2cm,height=4.2cm]{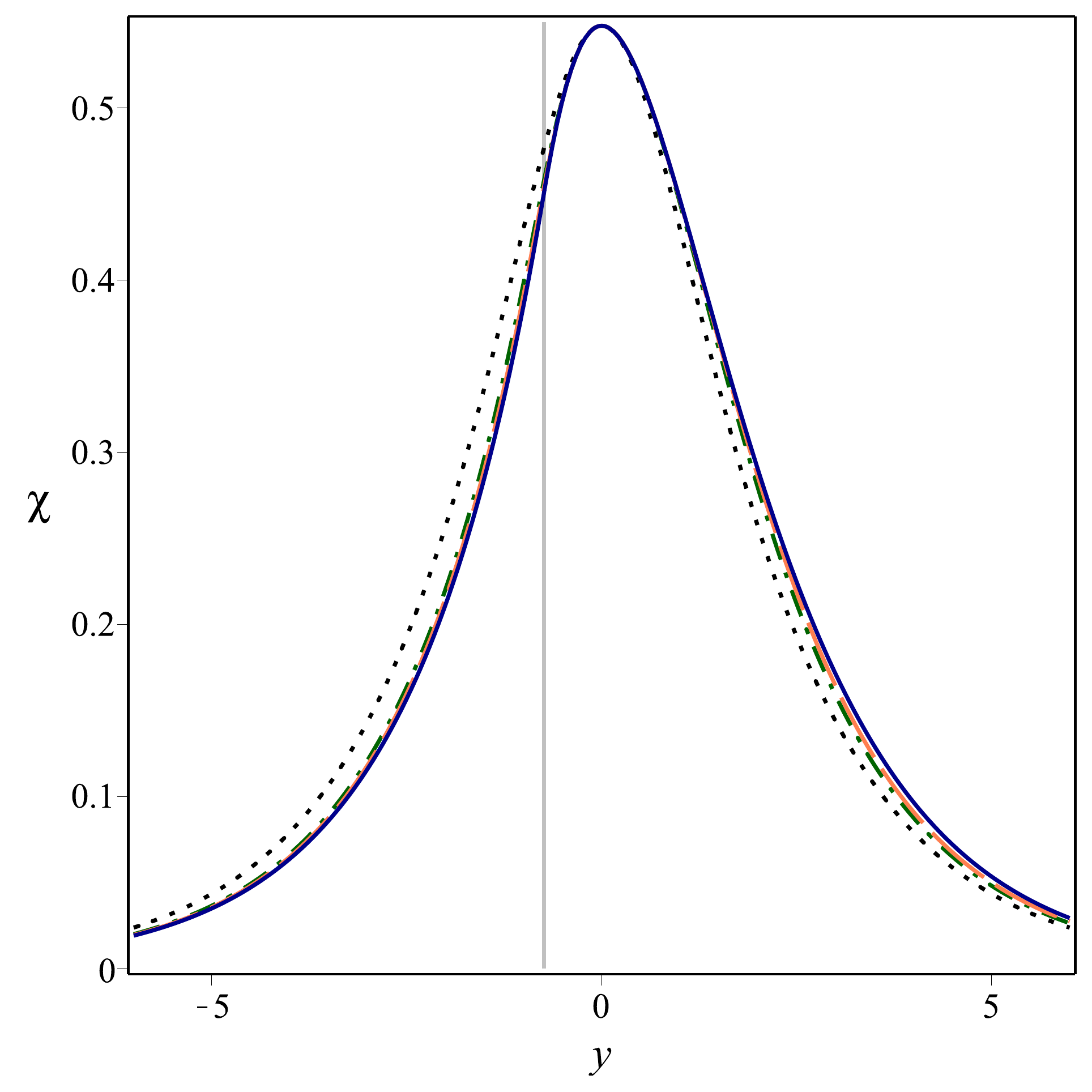}
\includegraphics[width=4.2cm,height=4.2cm]{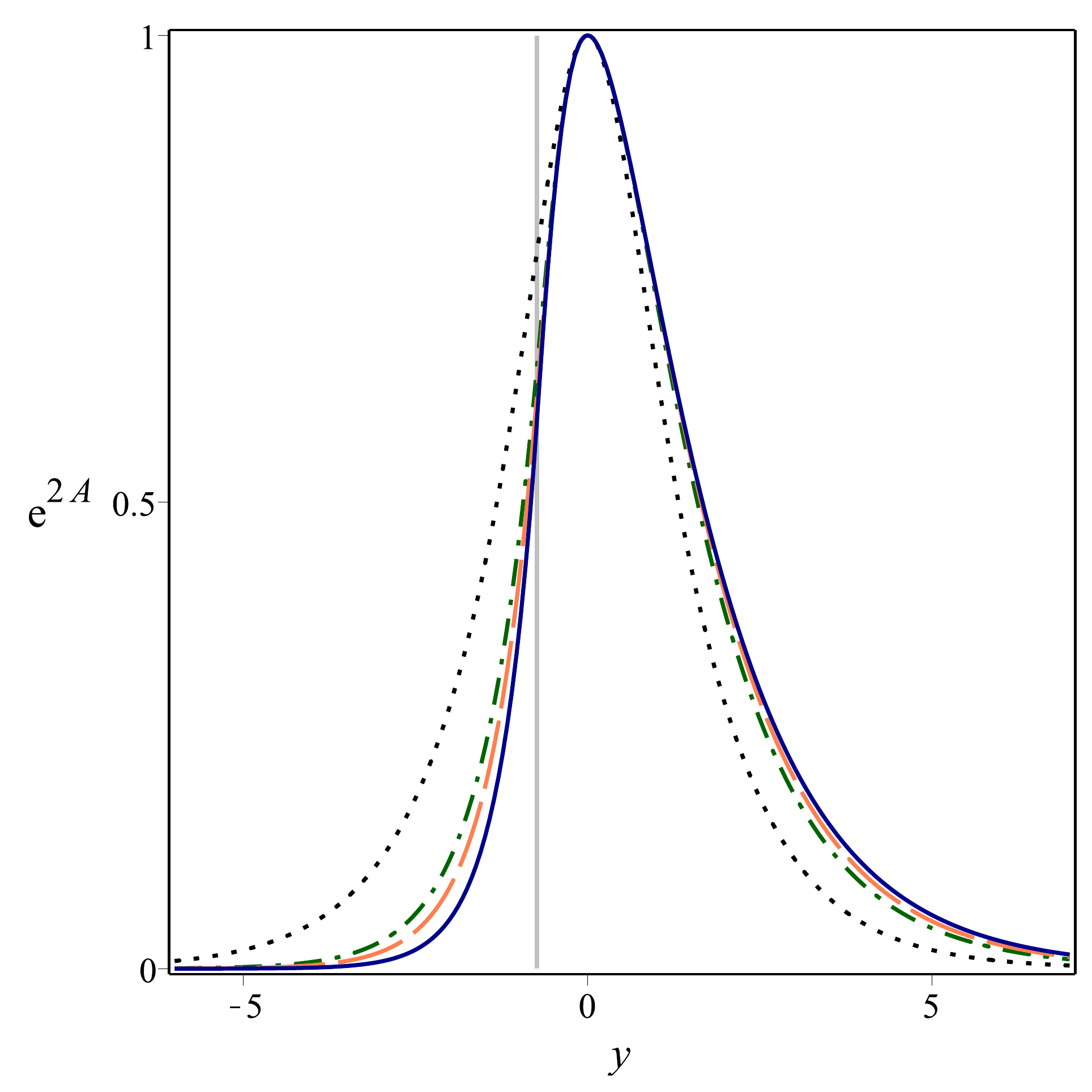}
\includegraphics[width=4.2cm,height=4.2cm]{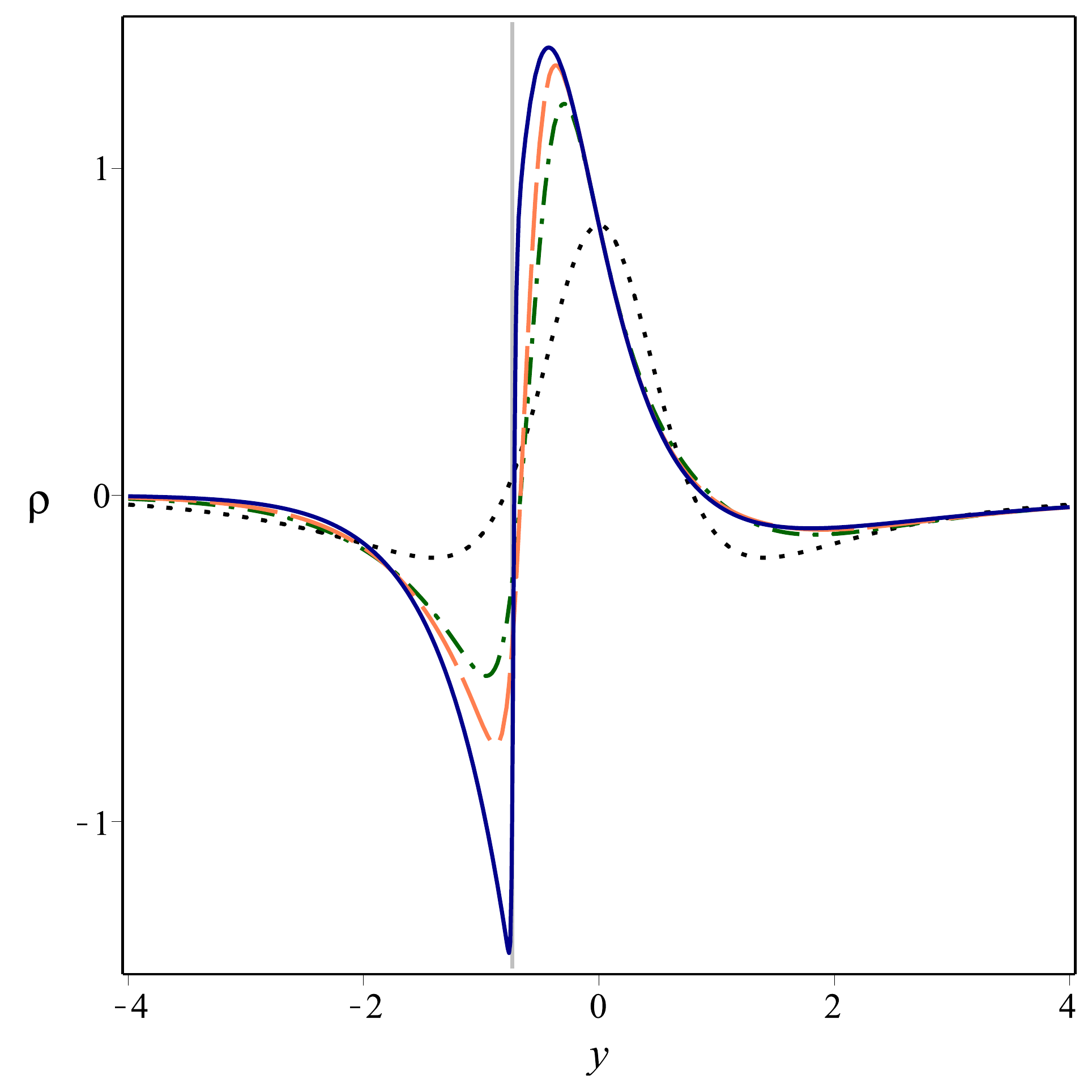}
\caption{In the top panel, field solutions $\phi(y)$ and $\chi(y)$. In the bottom panel, warp factor $\exp[2A(y)]$ and energy density $\rho(y)$. For the same values of $p$ presented in Fig. \ref{fig:5}, and $r=0.3$.}\label{fig:6}
\end{figure}

\begin{figure}
\includegraphics[width=4.2cm,height=4.2cm]{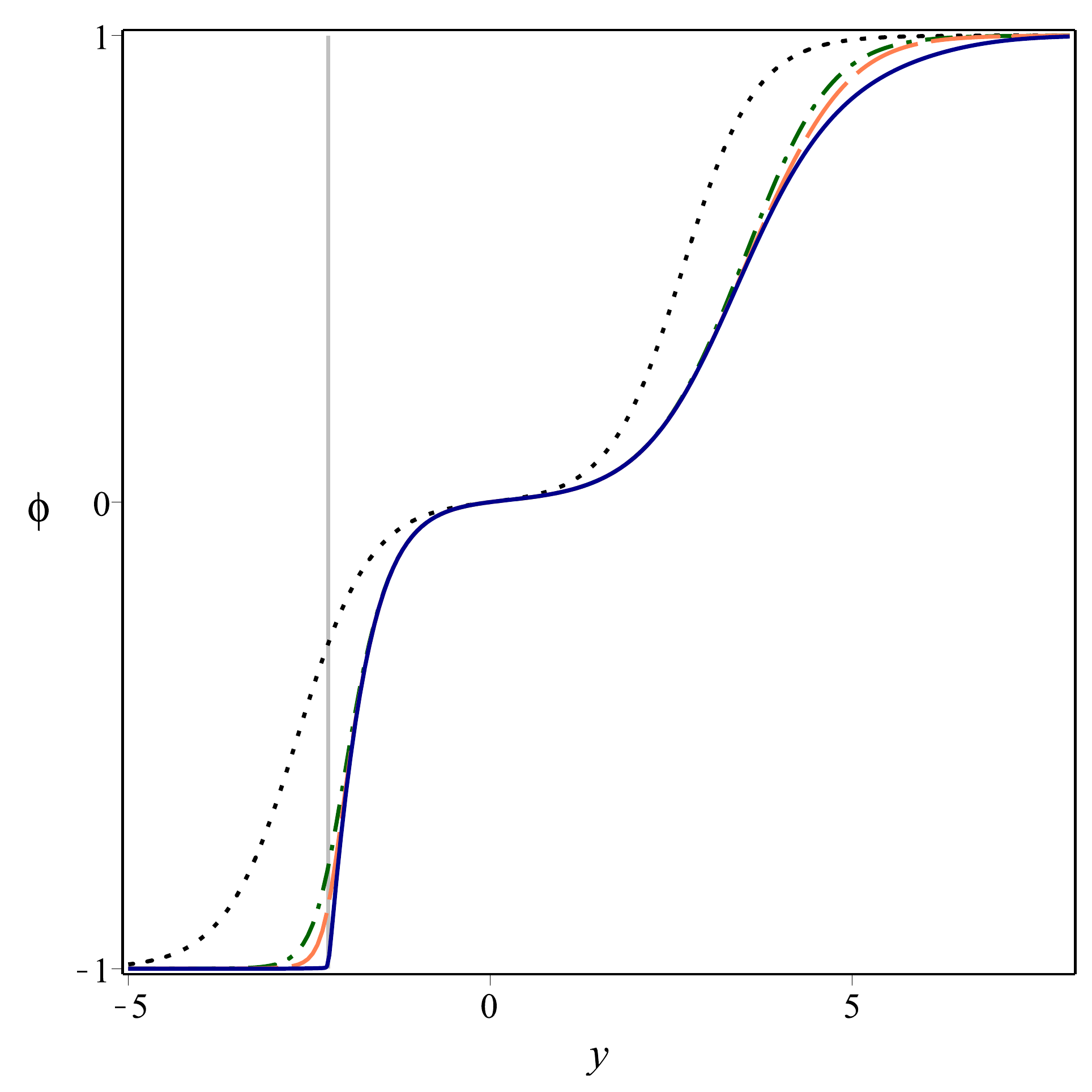}
\includegraphics[width=4.2cm,height=4.2cm]{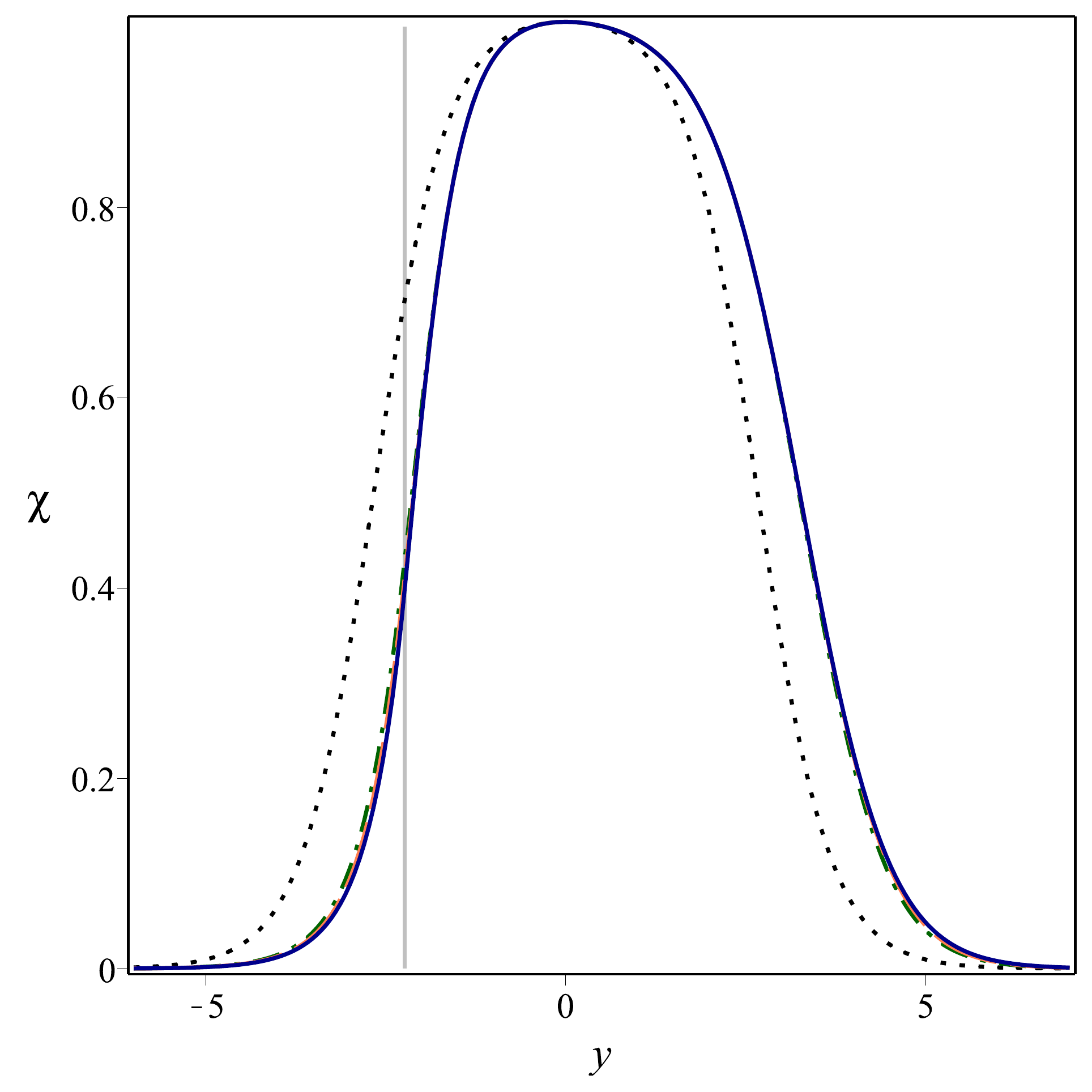}
\includegraphics[width=4.2cm,height=4.2cm]{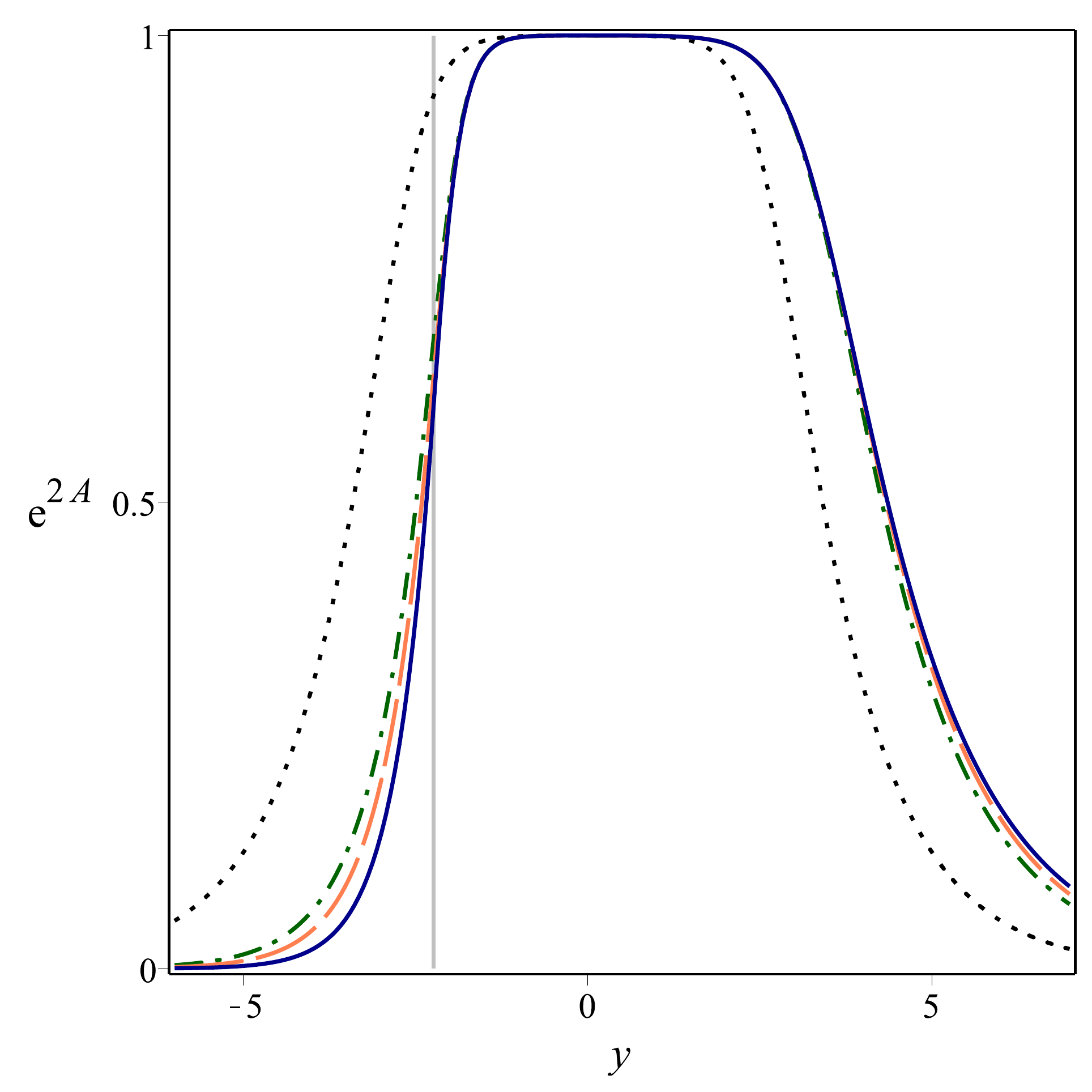}
\includegraphics[width=4.2cm,height=4.2cm]{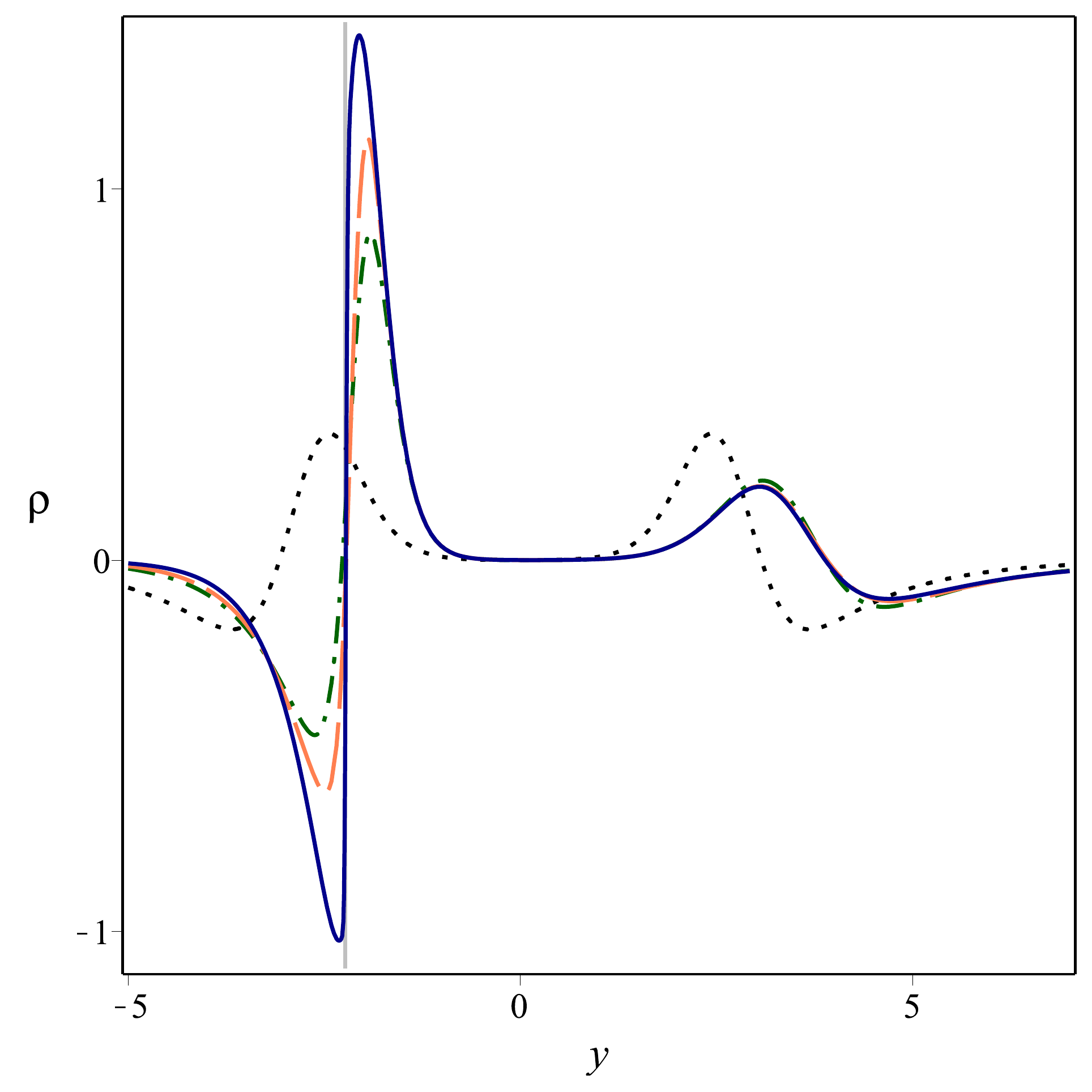}
\caption{In the top panel, we display the field solutions $\phi(y)$ and $\chi(y)$. In the bottom panel, we show the warp factor $\exp[2A(y)]$ and energy density $\rho(y)$. We consider the same values of $p$ used in Fig. \ref{fig:5}, and we take $r=0.99$.}\label{fig:7}
\end{figure}

\begin{figure}
\includegraphics[width=4.2cm,height=4.2cm]{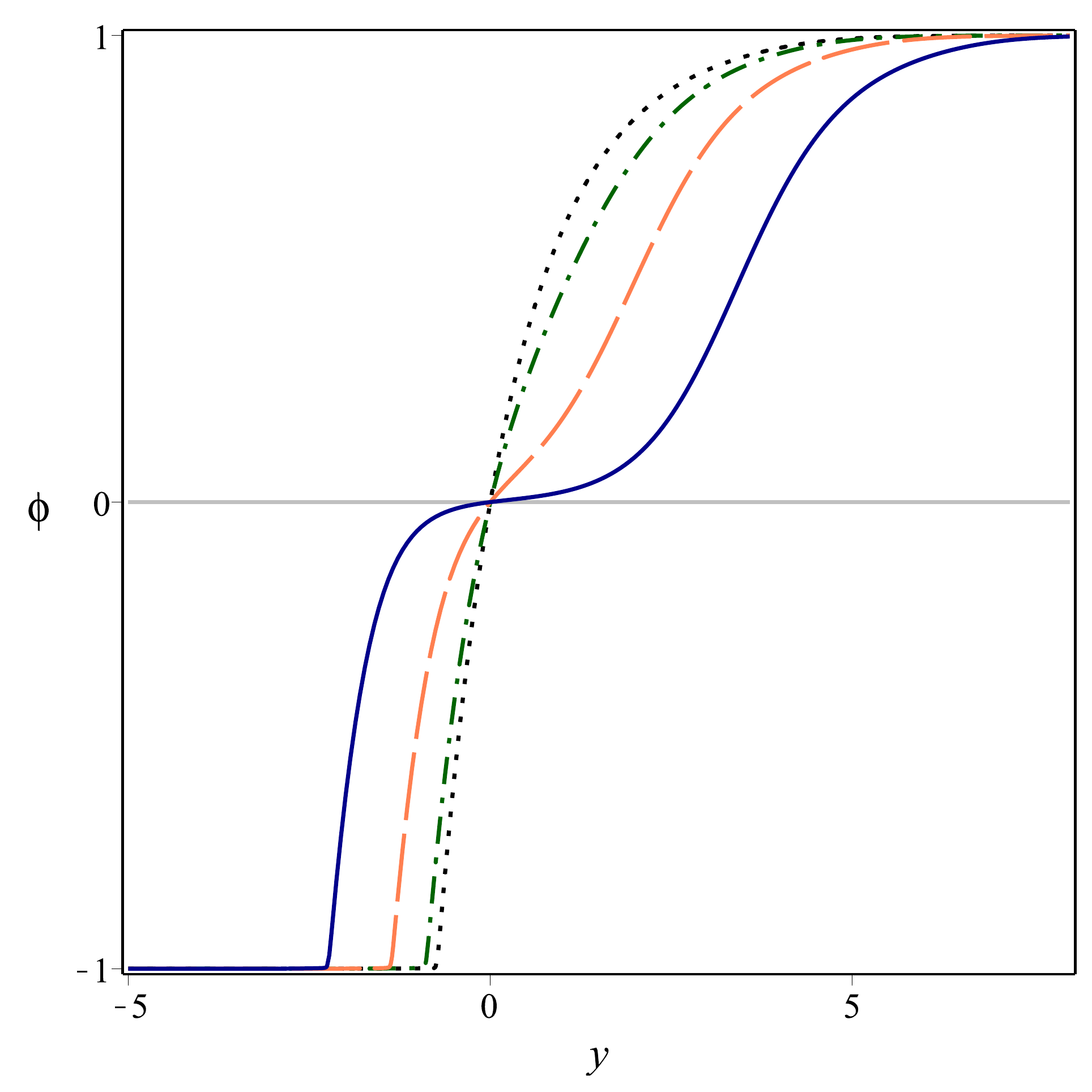}
\includegraphics[width=4.2cm,height=4.2cm]{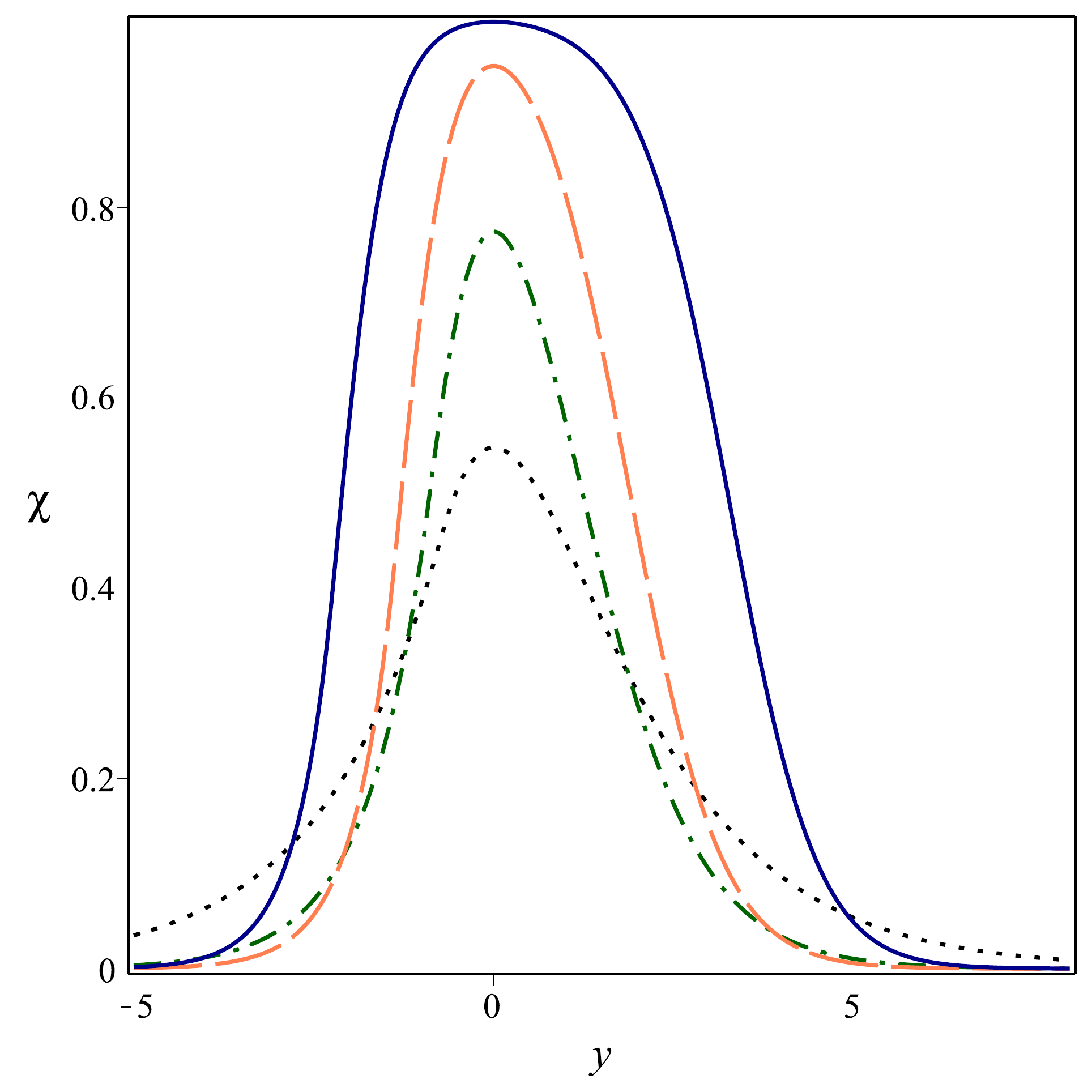}
\includegraphics[width=4.2cm,height=4.2cm]{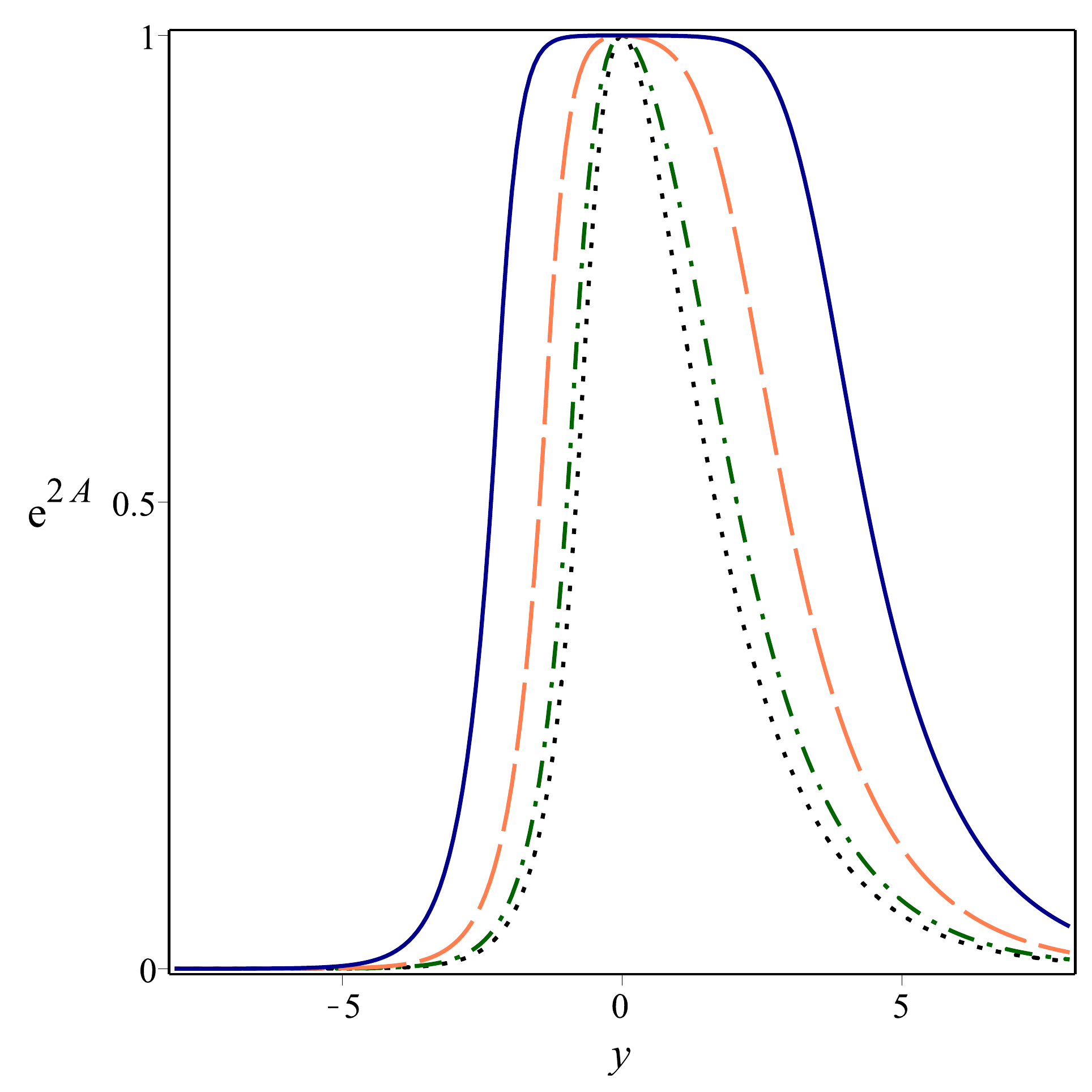}
\includegraphics[width=4.2cm,height=4.2cm]{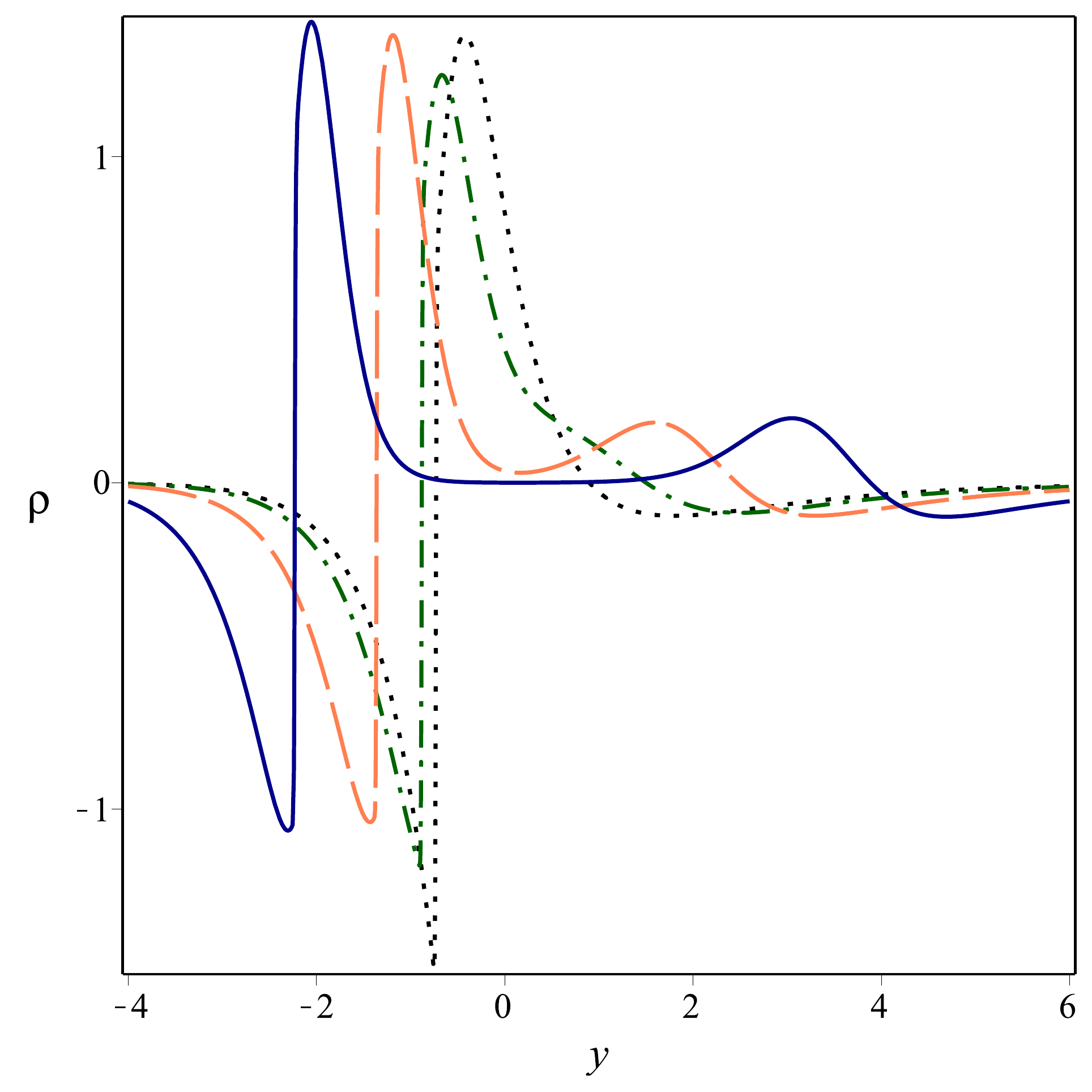}
\caption{In the top panel, we display the field solutions $\phi(y)$ and $\chi(y)$. In the bottom panel, we show the warp factor $\exp[2A(y)]$ and energy density $\rho(y)$. We consider $p=45$, and take $r=0.3, 0.6, 0.9, 0.99$, depicted with dotted (black), dot-dashed (green), dashed (red), and solid (blue) lines, respectively.}\label{fig:8}
\end{figure}

Now, we introduce a second type of model which supports an asymmetric hybrid brane behavior with internal structure. The model has the following superpotential 
\be\label{WW}
W(\phi,\chi)=\phi -\frac{\phi^2}{2}+\frac{\phi^{p+1}}{p+1}- \frac{\phi^{p+2}}{p+2}-r\phi\chi^2,
\ee
where $p$ is an odd integer, $p = 1, 3, 5\cdots$. It reproduces the Bloch brane model \cite{BB}, for $p=1$. The potential for the brane has the profiles  
\ben \label{pp01}
V(\phi,0)&=&\frac{1}{2} (1-\phi)^2(1+\phi^p)^2\nonumber \\
&&-\frac{4}{3}\phi^2\left(1-\frac{\phi}{2}+\frac{\phi^p}{p+1}-\frac{\phi^{p+1}}{p+2}\right)^2;\\
\label{pq01}
V(0,\chi)&=&\frac{1}{2}(1-r\chi^{2})^2.
\een
This model is represented in Fig. \ref{fig:5}. It satisfies first-order equations and admits interesting kinklike and lumplike solutions, depending on the value of the parameter $p$, as we now investigate. The study is more complicated in this case, so we have developed numerical calculations which we show in Figs.~\ref{fig:6}, \ref{fig:7} and \ref{fig:8}. These figures have similar meaning to Figs. \ref{fig:2}, \ref{fig:3} and \ref{fig:4}, that appeared before for the model 1, already explained in the previous subsection.

The results show that for large $p$, the brane is asymmetric and hybrid, since it behaves as a thin or
thick brane, depending on location of the extra dimension $y$. In special, in Fig. \ref{fig:7} one shows that as $p$ increases to larger and larger values, the 2-kink solution becomes an asymmetric half-compact structure, differently from the previous model. Furthermore, the warp factor engenders an asymmetric profile and the energy density has an asymmetric and exotic behavior, although it also results on null total energy.

In Fig. \ref{fig:8} we indicate how internal structure appears in this hybrid brane scenario. The brane thickness increases as $r$ goes to unity. Furthermore, the increase in $r$ leads to appearance of an exotic asymmetric structure that behaves as a half-compact and a standard structure, which contributes to change the behavior of energy density, with the appearance of two asymmetric maxima (see Fig.~\ref{fig:8}, bottom panel, right), one related to the half-compact structure and the other to the standard solution. Although this behavior is obtained numerically, one can show the hybrid profile of the solution analytically. To make this point clear, we note from Fig.~\ref{fig:8} (top panel, left) that when
$\phi$ reaches the value $-1$, which appears for negative values of $y$, the field $\chi$ and the warp function behave as
\ben
 \chi(y)&=&c_{1}(r)\e^{-2r|y|},\label{chip} \\
A(y)&=&-|y|-\frac{c_{1}(r)^2}{6}\e^{-4r|y|}+c_{2}(r), \label{warpp}
\een
where $c_1(r)$ and $c_2(r)$ are integration constants, so the warp factor $e^{2A}$ cleary engenders the thin wall behavior. To see how this behavior appears, one firstly notes that for $\phi\to-1$ and $p\gg1$ (see Fig.~\ref{fig:8}, top panel, left) the function in Eq.~\eqref{WW} becomes $W=-3/2+r\chi^2$; thus, the first-order equation for $\chi$ becomes $\chi^\prime=2r\chi$ and we can write $\chi(y)$ as in the form shown in Eq.~\eqref{chip}. A similar investigation follows for the warp factor, with the first-order equation given in Eqs.~\eqref{eqq}; here we have $A^\prime=1-(2/3)r\chi^2$, which leads to the behavior shown in the above Eq.~\eqref{warpp}.

\subsection{Linear stability}

\begin{figure}
\includegraphics[width=4.2cm,height=4.5cm]{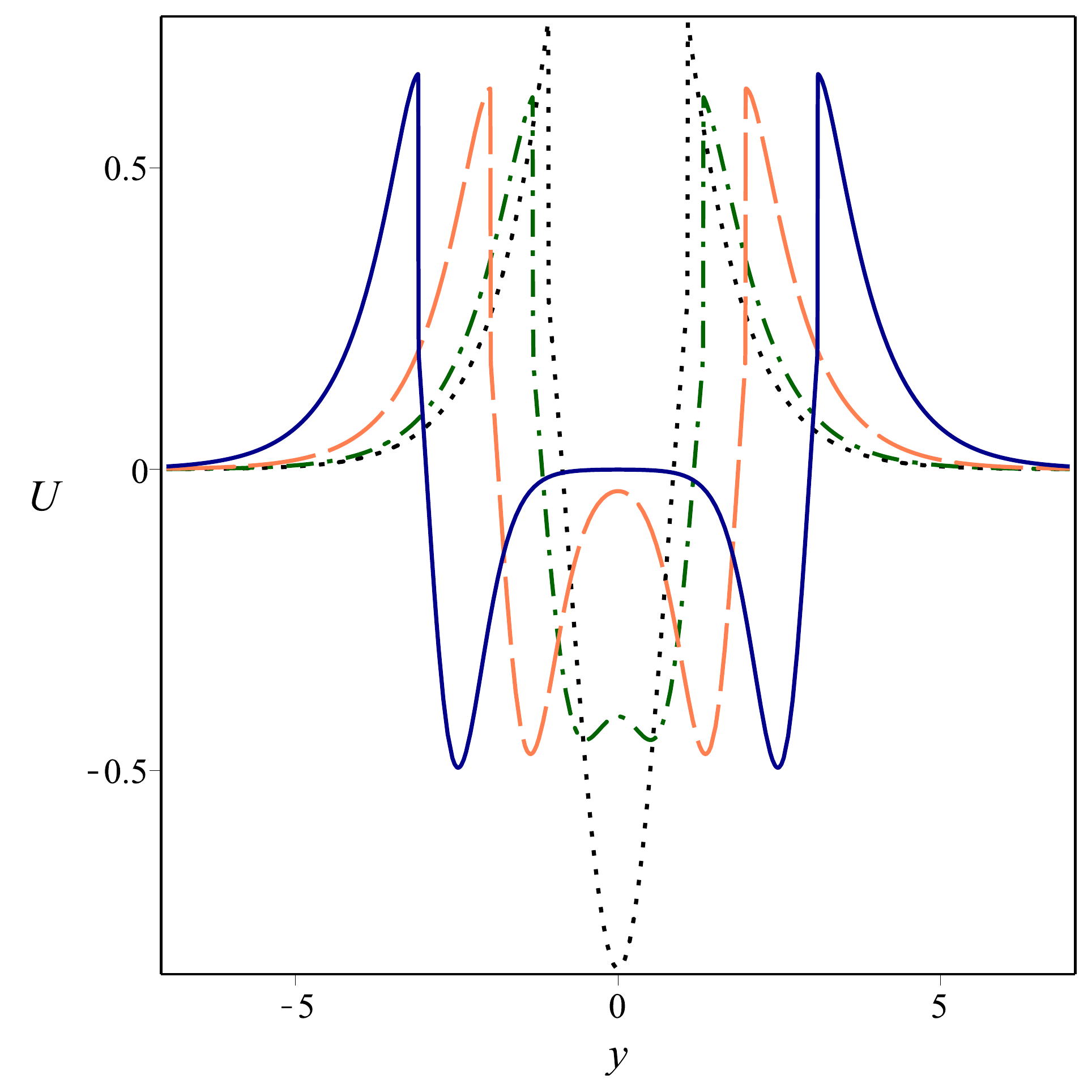}
\includegraphics[width=4.2cm,height=4.5cm]{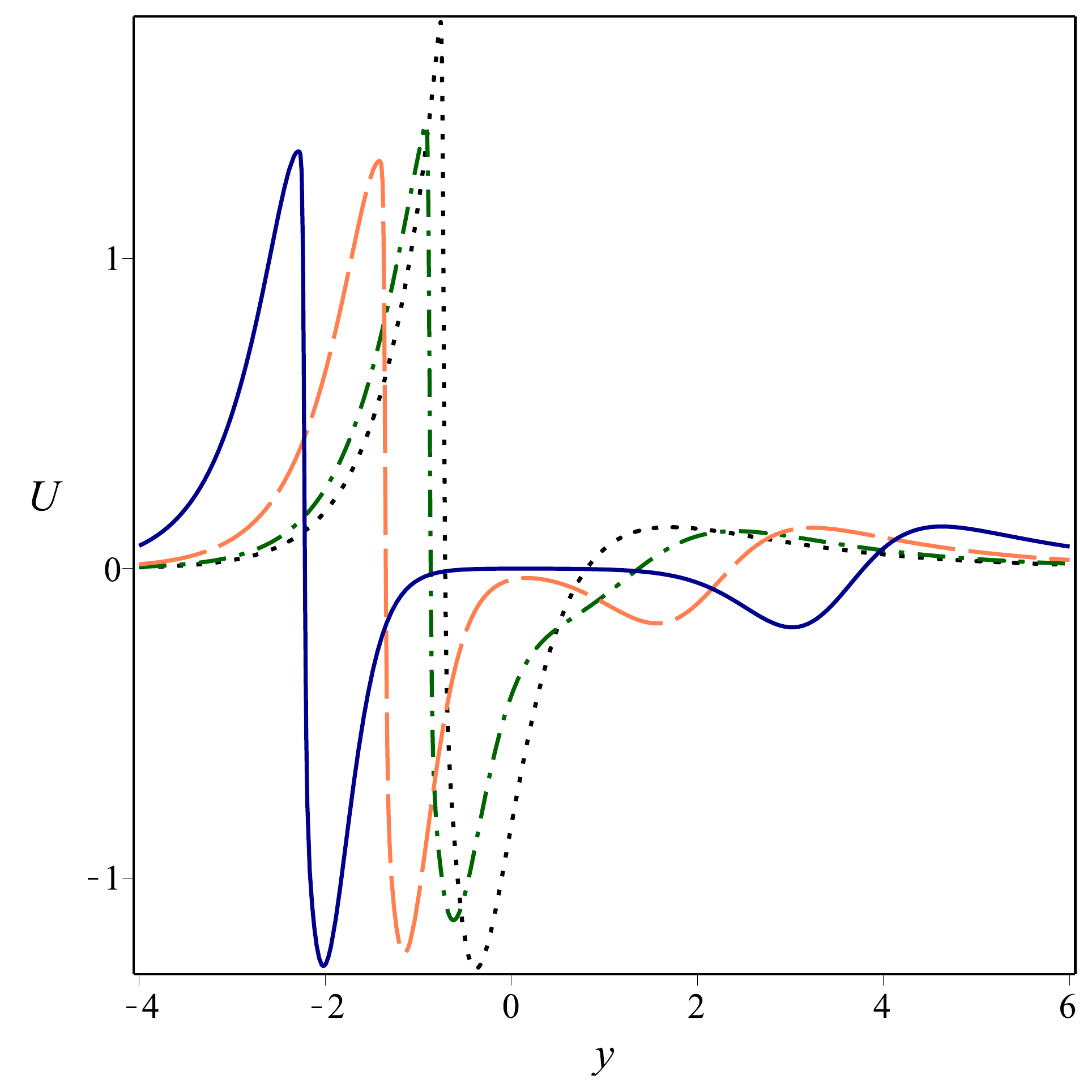}
\caption{Stability potential of the hybrid Bloch branes for $r=0.3, 0.6, 0.9, 0.99$, depicted with dotted (black), dot-dashed (green), dashed (red), and solid (blue) lines, respectively. In left panel, we present the stability potential for the model 1, while for the model 2 it is shown in right panel.}\label{fig:9}
\end{figure}

We have analyzed the linear stability of the gravity sector of the two new scenarios, for the two models under consideration. In particular, the two stability potentials that appear from the two models are displayed in Fig. \ref{fig:9}. They support the zero mode and no other bound state, and inform us that the two scenarios are stable under small fluctuations of the metric.

\section{Comments and conclusions}\label{sec-com}

In this work we explored the presence of internal structure in hybrid braneworld scenarios. We
proposed models described by two real scalar fields coupled to gravity in $(4,1)$ dimensions with warped geometry, which generate the so-called Bloch brane. The structure inside the brane is controlled by a coupling constant that determines the strength of interaction between the two fields. With these informations in hands, we suggested a mechanism that transforms thick Bloch branes into hybrid Bloch branes, through the introduction of a new parameter. Solutions for the source fields, warp factor, energy density, and stability potential are presented. Moreover, we addressed one symmetric situation, which develops a splitting of the energy density of the brane, and another one, asymmetric, which provides similar but now asymmetric behavior. 

The results open new possibilities, for investigations involving degeneracy and criticality on the symmetric hybrid Bloch brane, taking into account variations in the orbit \eqref{orbt} to test existence of new field solutions, in a way similar to the studies implemented in
Refs.~\cite{Dutra,Dutra2005,Correa}. Other issues of interest demand further investigations, in particular on how fermions may be localized on these branes, using different types of coupling between spinors and scalar fields, following the lines of Refs.~ \cite{Almeida,Castro,Correa,FL1,FL2,Zhao}. An interesting issue here concerns fermion localization in the second model, since it leads to an asymmetric brane, and this may avoid fermion localization under standard coupling, but this may be solved with the recent new possibility \cite{FL3}. Also, one can study the presence of resonances in both the symmetric and the asymmetric cases \cite{Almeida,Cruz2014,Xie}. It is also of interest to study how the hybrid Bloch brane scenario changes as one modifies the internal geometry of the brane \cite{FN,dsads}, considering bent branes of the de Sitter or anti de Sitter type. Another study of interest concerns modifications of the Newton's law, that is, the investigation of how the parameters $n$ and $p$ that are used to build the hybrid Bloch brane scenarios may modify gravity at the nonrelativistic limit. 

The possibility that our universe is described by a braneworld has also led to the study of braneworld cosmology and the domain-wall/brane-cosmology correspondence, so it is also of interest to study how the hybrid brane concept investigated in \cite{FC} and in the current work can be embedded into the braneworld cosmology scenario, as it appears, for instance, in Refs.~\cite{A,B,C}. 

Inspired by the AdS/CFT correspondence, in recent years the new concept of holographic cosmology has also being explored, relying on the study of a class of $(3,1)$ time-dependent metrics induced on slices of the $(4,1)$ dimensional asymptotic AdS$_5$ bulk; for more on this see, e.g., Ref.~\cite{T,HC} and references therein. An issue of interest appears to be the extension of the hybrid brane concept to such string-inspired holographic studies. The challenge to embed the hybrid brane into a bent brane \cite{FN,dsads} is also of current interest in connection to the construction of curved brane with regular support, as recently investigated in Refs.~\cite{A1}. These and other related issues are now under investigation, and we hope to report on them in the near future.

\section*{Acknowledgments}

This work is partially supported by the Brazilian agency CNPq. DB acknowledges support from the CNPq projects 455931/2014-3 and 306614/2014-6, EEML acknowledges support from the CNPq project 160019/2013-3, and LL acknowledges support from the CNPq projects 307111/2013-0 and
447643/2014-2.


\begin{thebibliography}{99}
\bibitem{Rubakov} V.A. Rubakov and M.E. Shaposhnikov, Phys. Lett. B {\bf125}, 139 (1983).
\bibitem{CCP} S. Weinberg, Rev. Mod. Phys. {\bf61}, 1 (1989); S. M. Carroll, Living Rev. Rel. {\bf4}, 1 (2001); P.J.E. Peebles and B. Ratra, Rev. Mod. Phys. {\bf75}, 559 (2003).
\bibitem{CD} N. Arkani-Hamed, S. Dimopoulos and G. Dvali, Phys. Lett. B {\bf429}, 263 (1998); I. Antoniadis, N. Arkani-Hamed, S. Dimopoulos and G. Dvali, Phys. Lett. B {\bf436} 257 (1998). 
\bibitem{RA} L. Randall and R. Sundrum, Phys. Rev. Lett. {\bf83}, 3370 (1999).
\bibitem{RA2} L. Randall and R. Sundrum, Phys. Rev. Lett. {\bf83}, 4690 (1999).
\bibitem{Arkani} N. Arkani-Hamed, S. Dimopoulos, N. Kaloper and R. Sundrum, Phys. Lett. B {\bf480}, 193 (2000).
\bibitem{FD} O. DeWolfe, D.Z. Freedman, S.S. Gubser and A. Karch, Phys. Rev. D {\bf62}, 046008 (2000).
\bibitem{Grem} M. Gremm, Phys. Lett. B {\bf478}, 434 (2000).
\bibitem{Csaki} C. Csaki, J. Erlich, T.J. Hollowood and Y. Shirman, Nucl. Phys. D, {\bf581}, 309 (2000).
\bibitem{BF} D. Bazeia, C. Furtado and A.R. Gomes, JCAP {\bf0402}, 002 (2004).
\bibitem{BB} D. Bazeia and A.R. Gomes, JHEP {\bf405}, 012 (2004).
\bibitem{Dutra2005} A.S. Dutra, Phys. Lett. B {\bf626}, 249 (2005).
\bibitem{Dutra} A.S. Dutra, A.C.A. Faria and M. Hott, Phys. Rev. D {\bf78}, 043526 (2008). 
\bibitem{Almeida} C.A.S. Almeida, M.M. Ferreira, A.R. Gomes and R. Casana,  Phys. Rev. D {\bf79}, 125022 (2009).
\bibitem{Zhao} Z-H Zhao, Y-X Liu and H-T Li, Class. Quant. Grav. {\bf27}, 185001 (2010).
\bibitem{Castro} L.B. Castro, 	Phys. Rev. D {\bf83}, 045002 (2011).
\bibitem{Correa} R.A.C. Correa, A.S. Dutra and M.B. Hott, Class. Quant. Grav. {\bf28}, 155012 (2011).
\bibitem{FL1}A.E.R. Chumbes, A.E.O. Vasquez, and M.B. Hott, Phys. Rev. D {\bf83}, 105010 (2011).
\bibitem{Xie} Q-Y Xie, J. Yang and L. Zhao, Phys. Rev. D {\bf88}, 105014 (2013).
\bibitem{Cruz} W.T. Cruz, Aristeu R.P. Lima and C.A.S. Almeida, Phys. Rev. D {\bf87}, 045018 (2013); 
W. T. Cruz, R. V. Maluf and C. A. S. Almeida, Eur. Phys. J. C {\bf73}, 2523 (2013).
\bibitem{FL2}Y.-X. Liu, Z.-G. Xu, F.-W. Chen, and S.-W. Wei, Phys. Rev. D {\bf89}, 086001 (2014).
\bibitem{Cruz2014} W.T. Cruz, L.J.S. Sousa, R.V. Maluf and C.A.S. Almeida, Phys. Lett. B {\bf730}, 314 (2014).
\bibitem{FC} D. Bazeia, L. Losano, M.A. Marques and R. Menezes, Phys. Lett. B {\bf736}, 515 (2014).
\bibitem{CE} D.F.S. Veras, W.T. Cruz, R.V. Maluf and C.A.S. Almeida, Phys. Lett. B {\bf754}, 201 (2016).
\bibitem{AHB} D. Bazeia, M. A. Marques and R. Menezes, Phys. Rev. D {\bf92}, 084058 (2015).
\bibitem{FN}D.Z. Freedman, C. N\'u\~nez, M. Schnabl, and K. Skenderis, Phys. Rev. D {\bf69}, 104027 (2004).
\bibitem{dsads}V.I. Afonso, D. Bazeia, and L. Losano, Phys. Lett. B {\bf634}, 526 (2006); 
D. Bazeia, F.A. Brito, and L. Losano, JHEP {\bf0611}, 064 (2006).
\bibitem{A}M. Cvetic and H. H. Soleng, Phys. Rep. {\bf282}, 159 (1997).
\bibitem{B}K. Skenderis and P. K. Townsend, Phys. Rev. Lett. {\bf96}, 191301 (2006).
\bibitem{C}D. Bazeia, F.A. Brito, F.G. Costa, Phys. Lett. B {\bf661}, 179 (2008). 
\bibitem{T}P.S. Apostolopoulos, G. Siopsis and N. Tetradis, Phys. Rev. Lett. {\bf102}, 151301 (2009).
\bibitem{HC}N. Bilic, Phys. Rev. D {\bf93}, 066010 (2016). 
\bibitem{BNRT}D. Bazeia, M.J. dos Santos and R. Ribeiro, Phys. Lett. A {\bf208}, 84 (1995); D. Bazeia and M.M. Santos, Phys. Lett. A {\bf217},
28  (1996); D. Bazeia, J.R.S. Nascimento, R.F. Ribeiro, and D. Toledo, J. Phys. A {\bf30}, 8157 (1997).
\bibitem{TK} D. Bazeia, J. Menezes and R. Menezes, Phys. Rev. Lett. {\bf91}, 241601 (2003).
\bibitem{AC} A. Campos, Phys. Rev. Lett. {\bf88}, 141602 (2002); Z-H. Zhao, Y-X. Liu, Y-Q. Wang and H-T. Li, JHEP {\bf1106},  045 (2011).
\bibitem{FL3}Y.-Y. Li, Y.-P. Zhang, W.-D. Guo, and Y.-X. Liu, {\it Fermion localization mechanism with derivative geometrical coupling on branes}. arXiv:1701.02429. 
\bibitem{A1}I. Antoniadis, S. Cotsakis, I. Klaoudatou, Eur. Phys. J. C. {\bf76}, 511 (2016).
\end{thebibliography}
\end{document}